\documentclass{pasa}%

\usepackage{times}
\usepackage{graphicx,latexsym, amssymb, amscd, psfrag}
\usepackage{color}
\usepackage{subfig}
\usepackage{morefloats,rotating,float}
\usepackage{gensymb}
\usepackage{multirow,array}
\usepackage{adjustbox}
\usepackage{gensymb}
\usepackage{mathtools}

\title[Coma UVIT]{Deepest far ultraviolet view of a central field in the Coma cluster by {\it AstroSat} UVIT}

\author[Mahajan et al.]{Smriti Mahajan$^1$\thanks{Corresponding author; Email: smritimahajan@iisermohali.ac.in}, Kulinder Pal Singh$^1$, Joseph E. Postma$^2$, Kala G. Pradeep$^1$, Koshy George$^3$, Patrick Côté$^4$ 
\affil{$^1$Department of Physical Sciences, Indian Institute for Science Education and Research Mohali- IISERM, 
 Knowledge City, Manauli, 140306, Punjab, India }%
\affil{$^2$ Department of Physics and Astronomy, University of Calgary, 2500 University Dr NW, Calgary, Alberta T2N 1N4, Canada }
\affil{$^3$ Ludwig-Maximilians-Universit{\"a}t, Scheinerstr. 1, 81679, Munich, Germany }
\affil{$^4$ Herzberg Astronomy and Astrophysics Research Centre, National Research Council of Canada, 5071 W. Saanich Road, Victoria, BC V9E 2E7, Canada}
}%

\def\arcsec{{^{\prime\prime}}}
\def\g{{\it GALEX }}
\def\a{{\it AstroSat }} 
\definecolor{grey}{rgb}{0.5,0.6,0.7}

\definecolor{amber}{rgb}{1.0,0.49,0.0}

\newcommand{\ha}{{H$\alpha$}}

\jid{PASA}
\doi{10.1017/pas.\the\year.xxx}
\jyear{\the\year}

\usepackage{aas_macros}
\usepackage{hyperref} 
\hypersetup{colorlinks,citecolor=blue,linkcolor=blue,urlcolor=blue}

\hypersetup{draft}

\begin{document}

\begin{frontmatter}
\maketitle

\begin{abstract}
 We present analysis of the far ultraviolet ({\it FUV}) emission of sources in the central region of the Coma cluster ($z=0.023$) using the data taken by the UVIT aboard
 the multi-wavelength satellite mission {\it AstroSat}. We find a good correlation between the UVIT {\it FUV}  flux and the fluxes in both wavebands of the {\it Galex} mission, 
 for the common sources. We detect stars and galaxies, amongst which the brightest ($r \lesssim 17$ mag) galaxies in the field of view are mostly 
 members of the Coma cluster. We also detect three quasars ($z = 0.38, 0.51, 2.31$), one of which is likely the farthest object observed by the UVIT so far. 
 In almost all the optical and UV colour-colour and colour-magnitude planes explored in this work, the Coma galaxies, other galaxies and 
 bright stars could be separately identified, but the fainter stars and quasars often coincide with the faint galaxies. We have also investigated galaxies with unusual FUV 
 morphology which are likely to be galaxies experiencing ram-pressure stripping in the cluster.  Amongst others, two confirmed cluster members which were not 
 investigated in the literature earlier, have been found to show unusual FUV emission. All the distorted sources are likely to have fallen into the cluster recently, and hence have 
 not virialised yet. A subset of our data have optical spectroscopic information available from the archives. For these sources ($\sim 10\%$ of the sample), we find that 17 galaxies 
 identify as star-forming, 18 as composite and 13 as host galaxies for active galactic nuclei, respectively on the emission-line diagnostic diagram. 
\end{abstract}

\begin{keywords}
galaxies: evolution; galaxies: fundamental parameters; galaxies: star formation
\end{keywords}
\end{frontmatter}

 \section{Introduction}
 \label{intro}

 The broadband ultraviolet (UV) light (1000--3000 \AA) from galaxies offers a unique window into their evolution. The far UV light (FUV; 1000--2000 \AA) in normal
 star-forming galaxies is produced by intermediate-mass (2-5 $M_{\odot}$) short-lived stars ($\leq 1$ Gyr), and hence is a good tracer of the star formation rate 
 \citep[SFR;][]{kennicutt98,mahajan19}. On the other hand, low-mass helium burning stars which evolve through the horizontal branch produce FUV emission in 
 the more evolved elliptical galaxies ($> 8-10$ Gyr old), resulting in a UV upturn in the spectra of such galaxies at wavelengths shorter than 2000 \AA. In either case, the FUV emission can 
 therefore be used as an age-dating tool to assess the evolution history for a variety of galaxies.
 
 The Indian multi-wavelength satellite \a launched in 2015 \citep{Si2014}  greatly surpassed its predecessors in observing the UV sky. The {\it Ultraviolet Imaging Telescope} 
 \citep[UVIT;][]{tandon17, Tan20} on board the \a offers the highest spatial resolution of $\lesssim 1.8^{\prime\prime}$ relative to similar missions such as the {\it Swift--UVOT, XMM-OM} and 
 {\it Galaxy evolution explorer} mission {\it Galex}.
 With a $28^{\prime}$ diameter field of view, UVIT is second only to {\it Galex} ($1.2\degree$), although the latter had a spatial resolution of $\sim 5^{\prime\prime}$.
 The large field of view together with the unprecedented resolution therefore makes UVIT a great tool for exploring the recent star formation history of galaxies in galaxy clusters.
 
 The Coma cluster ($\sim 100 h^{-1}$ Mpc, $z=0.023$) is one of the richest and the most well studied cluster of galaxies in the nearby Universe. Hence it is not 
 surprising that this cluster has extensive panchromatic coverage from X-rays to the radio continuum. Prior to the launch of the 
 {\it AstroSat} mission, the deepest UV observations ($\sim 26$ ks) of an off-centre field near the core of the Coma cluster were obtained by the 
 {\it Galex} mission \citep{hammer10}. This region lies $\sim 1\degree$ away from the 
 core, and is linked with the secondary peak in the X-ray emission from the Coma cluster \citep{finoguenov03, neumann03}, associated with the infalling galaxy group 
 NGC~4839 and several post-starburst galaxies \citep{poggianti04, mahajan10}. 
 
 Amongst others, the balloon-borne imaging FOCA instrument (2000 \AA) was used to conduct a UV survey of the core of the Coma cluster with a spatial resolution of 
 $\sim 20^{\prime\prime}$ which resulted in the first ever UV luminosity function for a cluster of galaxies \citep{donas91}. Follow up observations of several 
 other nearby clusters together with the Coma cluster showed that the UV luminosity functions are also well fitted by a Schechter function, and are similar to the ones 
 obtained for the general `field' \citep[e.g.][]{cortese03}.
 These observations indicate that the mechanisms responsible for quenching star formation of galaxies falling into clusters influence the 
 giant and dwarf galaxies in a similar fashion, therefore preserving the shape of the luminosity function, albeit with different normalization. 

 In this work we utilise the merits of the UVIT to explore a region $\sim 5.5^{\prime}$ away ($\sim 0.7$ Mpc at $z=0.023$) from the core 
 of the Coma cluster. The deviation from the core is 
 necessitated by the technical requirements of the UVIT instruments. In our knowledge, this is the first statistical study of galaxies based on UVIT data, and 
 hence a benchmark for the ones to follow from the arsenal of good quality data which is now becoming available. 

 In the next section we describe the UV and optical data used in this work, followed by 
 the properties of the detected UV sources in Sec.~\ref{props}. In Sec.~\ref{distorted} we discuss some individual sources with distorted UV morphology, followed by a 
 discussion of our findings in Sec.~\ref{discuss}. We finally conclude with the summary of our work in Sec.~\ref{summary}.   
 Throughout this work we use concordance $\Lambda$ cold dark matter cosmological model with $H_0 = 70$ km s$^{-1}$ Mpc$^{-1}$, $\Omega_\Lambda = 0.7$ 
 and $\Omega_m = 0.3$ to calculate distances and magnitudes. 
    
  \begin{figure*}
  \includegraphics[scale=0.85]{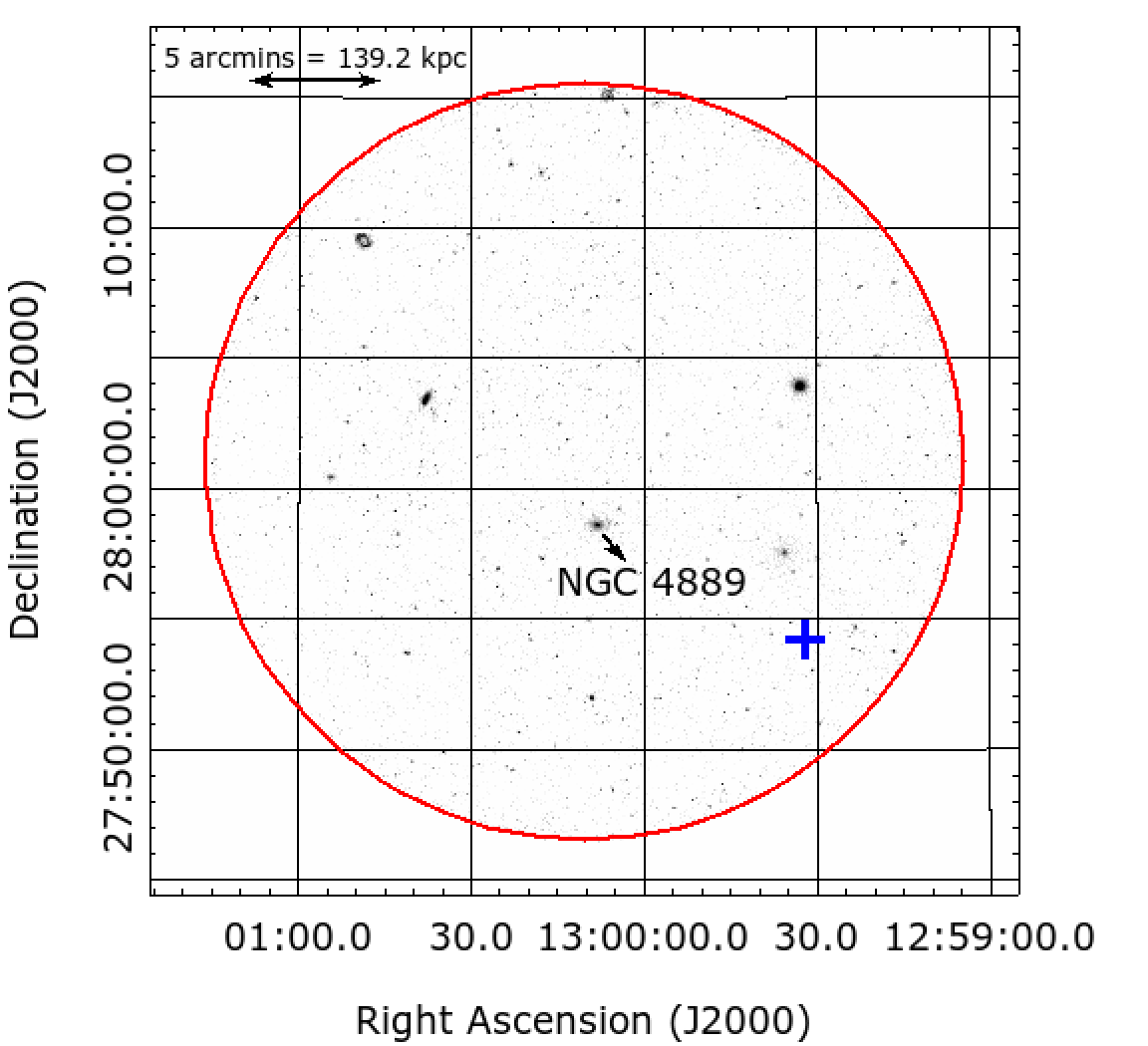}
 \caption{The UVIT field of view with the {\it red} circular region centred at $\alpha = 13:00:10.240$ and $\delta = 28:01:01.498$, and having a radius of 
 $14.5^\prime$ analysed in this paper. The {blue cross} denotes the x-ray centre of the Coma cluster \citep[$\alpha = 12:59:31.900, \delta = 27:54:10.000$;][]{rines03}. 
 The image is oriented such that North is up and east is on the left. The giant elliptical galaxy NGC 4889 is also shown for reference. } 
 \label{fov}
 \end{figure*}

 \begin{figure}
  \adjustbox{trim={.1\width} {.1\height} {0.1\width} {.1\height},clip}
  {\includegraphics[scale=0.4]{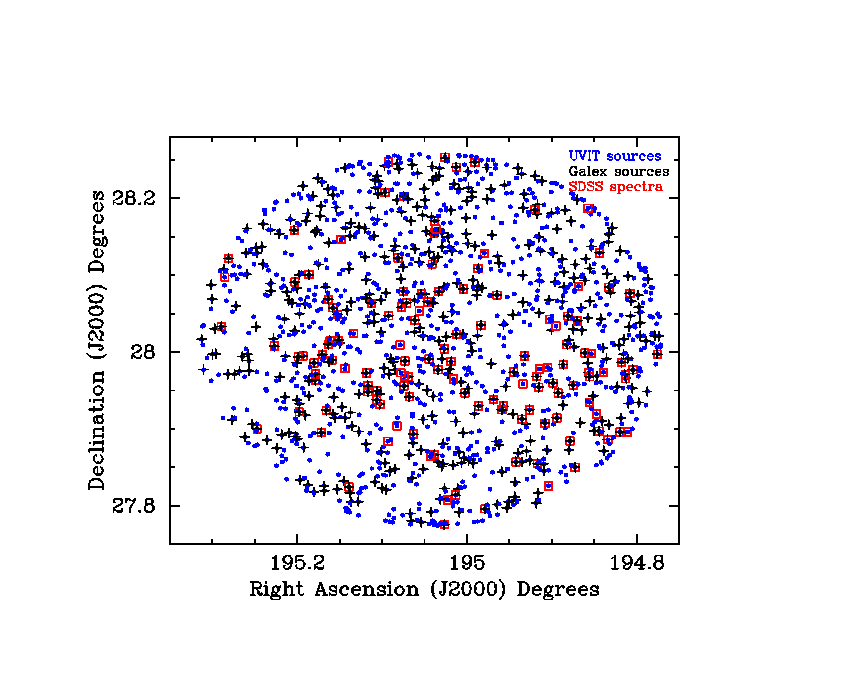}}
 \caption{This sky plot shows the distribution of 1308 sources {\it (blue points)} detected in the UVIT's FUV BaF$_2$ filter. The 134  
 sources with redshifts from the SDSS or NED are shown as {\it red squares}, and those also detected by \g are represented by {\it (black crosses)}, respectively. 
 This figure clearly shows that some UVIT detected sources are observed for the first time, while many UV sources lack redshift information.   } 
 \label{sky}
 \end{figure}

 \section{data}
 \label{data}

 \subsection{UV observations and data processing}
 \label{uvdata}
 
 In this paper we utilise the data from the UVIT on-board the \a \citep{kumar12}. The UVIT comprises two 35-cm Ritchey-Chr{\'e}tien telescopes 
 with one of them dedicated to the far ultraviolet (FUV; 1250--1830 \AA) and the other split between the near ultraviolet (NUV; 1900--3040 \AA) and the visible 
 (VIS; 3040--5500 \AA) channels, respectively. The raw data from the mission are processed by the Indian Space Science Data Centre (ISSDC) and the Level 1 
 as well as Level 2 (pre-processed events and images) from each instrument are provided to the users.
 
 The data used for the Coma cluster used in this work were taken as part of the guaranteed time observations in cycle 6 taken over 17 orbits of the {\it AstroSat} 
 (Observation ID: G06-077T01-9000001090; PI: K P Singh). Each orbit included simultaneous observations by VIS, FUV and X-ray instruments.
 The NUV channel could not be used due to the presence of a bright star in the field. 
 The FUV observations were done in the broad-band BaF$_2$ filter centred at 1541\AA\ \citep{tandon17, Tan20} with a total exposure time of 22669.5 seconds. 
  The final science image required manual source-tracking techniques to determine the drift, as detailed below, due to the field having only a single source bright enough for tracking in the VIS channel, as well as fainter sources which were interfered with by electronic artefacts in the VIS image data. The VIS data are not recommended for scientific analysis and are primarily used for tracking the spacecraft. Fig.~\ref{fov} shows the position of the UVIT field relative to the x-ray centre \citep[$\alpha = 13:00:10.240$ and $\delta = 28:01:01.498$;][]{rines03},
  along with the giant elliptical galaxy, NGC 4889.

 \a UVIT is a unique telescope because it oscillates its pointing on the orthogonal UVIT image axis at a rate of a few arcseconds every second. 
 These oscillations having an amplitude of 
 a few arcminutes protect the detector components from bright objects. The original image field for the UVIT is then recovered from the effects of these ``drifts" by de-shifting and 
 co-adding the count centroids of the detected sources, as a function of the pointing oscillations. For this work we processed the Level 1 FUV data with the CCDLAB software
  \citep{postma17}, following the procedure described in \citet{postma21}. The CCDLAB pipeline takes the L1 zip file as input and then performs various corrections on individual images 
  from each orbit after extraction. The CCDLAB pipeline automatically corrects for the translational drift using the VIS data along 
 with other procedures such as subtracting the flat field, detector corrections, and exposure array corrections, amongst others. 
 
 The corrected images are then co-aligned with manual intervention and stacked to create a deep field image. Following this, the automatic procedure for optimising the point source function (PSF) of the sources was initiated in order to remove the residual drift effects from the image, yielding 
  a final full width half maximum of $1.2^{\prime\prime}$ for this image. Finally, the world coordinate system (WCS) solution was obtained by mapping the pixel coordinates of sources in the
 image to the GAIA catalogue using the Astroquery utility \citep{ginsburg19} through CCDLAB, and its implementation of the automated WCS-solver as detailed in \citet{postma20}. The image was then de-rotated to match the sky coordinates, thus eliminating the effect of the telescope’s field rotation, and used for science. 
 
 The FUV source catalogue was created using the {\sc sextractor} software \citep{bertin96}. Sources were detected on a convolved version of the image (Gaussian kernel of 
 FWHM $ = 4.0$ pixels) in order to prevent shredding of larger galaxies. Other de-blending and detection parameters used by {\sc sextractor} were chosen after extensively 
 testing a wide range of values and visually testing the processed image. The background was detected by {\sc sextractor} in the ``AUTO" mode. The full set of {\sc sextractor} parameters used for this image are provided in Table~\ref{sextractor} in order to aid fellow UVIT users in processing similar imaging data. We detect 1308 sources in the UVIT BaF$_2$ filter (Fig.~\ref{fov}), which are analysed further in this paper. 
 
  \begin{table}
 \caption{{\sc Sextractor} (version v1.2b14) parameters used for the analysis of the UVIT  {\it FUV} image. }
 \begin{center}
 \begin{tabular}{ lc }     
 \hline
 Parameter & Value \\ \hline
THRESH\_TYPE  &     RELATIVE       \\   
DETECT\_MINAREA & 	20		 \\ 
DETECT\_THRESH	 & 2.0		 \\ 
ANALYSIS\_THRESH	 & 5.0	 \\ 	
FILTER		 & Y		 \\ 
FILTER\_NAME	 & gauss\_4.0\_7x7.conv \\ 	
DEBLEND\_NTHRESH & 	 30		 \\ 
DEBLEND\_MINCONT	 & 0.18	         \\ 
CLEAN		 & Y		 \\ 
CLEAN\_PARAM	 & 1.0		 \\ 
MASK\_TYPE	 & CORRECT  	 \\ 
PHOT\_APERTURES	 & 4.0		 \\ 
PHOT\_AUTOPARAMS	 & 0.3, 2.5       \\ 
PHOT\_FLUXFRAC  & 0.9 \\ 
PHOT\_PETROPARAMS  & 0.5, 2.5       \\ 
SATUR\_LEVEL	   & 55000            \\ 
GAIN		 & 1.0		 \\ 
PIXEL\_SCALE & 	0               \\ 
SEEING\_FWHM & 	1.5		 \\ 
STARNNW\_NAME	 & default.nnw	 \\ 
BACK\_TYPE       &  AUTO           \\ 
BACK\_SIZE	 & 8                \\ 
BACK\_FILTERSIZE	  & 5		 \\ 
BACKPHOTO\_TYPE	 & GLOBAL \\ 	         
BACKPHOTO\_THICK & 24	 \\ 	
 \hline
 \end{tabular}
 \end{center}
 \label{sextractor}  
 \end{table}
  
\begin{figure}
 \adjustbox{trim={.1\width} {.1\height} {0.1\width} {.1\height},clip}
 {\includegraphics[scale=0.39]{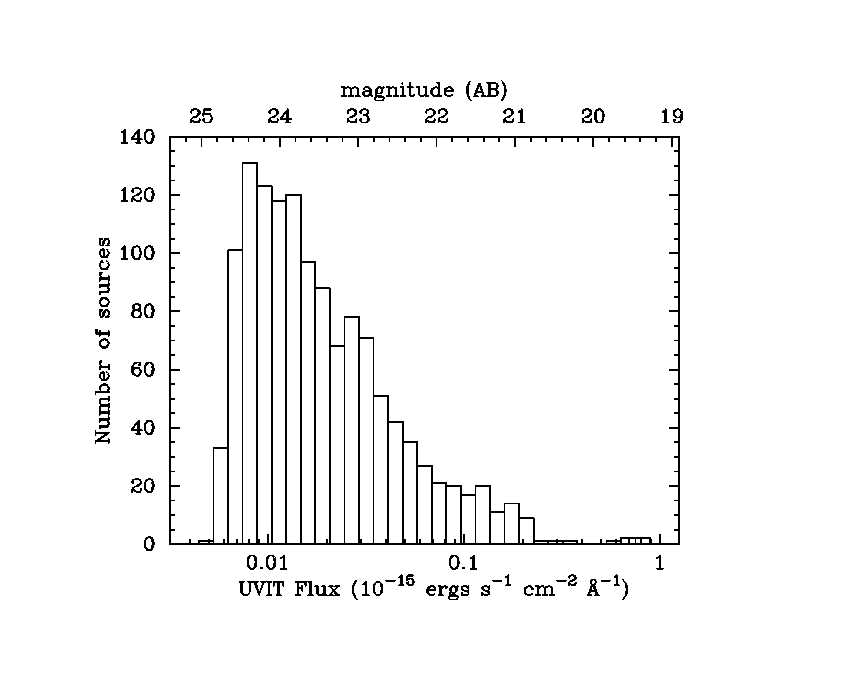}}
 \caption{The distribution of UVIT FUV flux corrected for the Milkyway extinction for all the 1308 sources analysed in this work. The top axis shows the AB magnitudes for the same. } 
 \label{flux}
 \end{figure}

 The integrated counts for all the 1308 sources identified in this image were normalized by the total integration time, and then converted to fluxes using the conversion factors
 provided by \citet{Tan20}. 
 The measured flux was corrected for galactic extinction assuming a Milky way-like extinction curve \citep{calzetti00}. Following \citet{bianchi11} we assume 
 $A_{FUV}=8.06 E(B-V)$. The $E(B-V)$ values are obtained for positions of all the identified sources from \url{https://irsa.ipac.caltech.edu/applications/DUST/}. This 
 resource provided the \citet{schlegel98} and \citet{schlafly11} estimates for reddening excess, of which we employed the latter estimate of $E(B-V)$ in this work. 
 We however note that 
 both estimates correlate very well for the sample studied in this work. The reddening excesses for these Coma cluster data are in the range  $0.0067 \leq E(B-V) \leq 0.0112$ with a median at $0.0090$ mag.  
 The distribution of extinction-corrected fluxes and the corresponding FUV magnitude for all the UVIT sources are shown in Fig.~\ref{flux}. This figure shows that our data
  spans almost three orders of magnitude in FUV flux, and a range of more than five magnitudes for the sources detected by the UVIT.  
    
 \subsection{Optical data}
 \label{opt}
 
  \begin{figure}
  \adjustbox{trim={.1\width} {.1\height} {0.1\width} {.1\height},clip}
 {\includegraphics[scale=0.39]{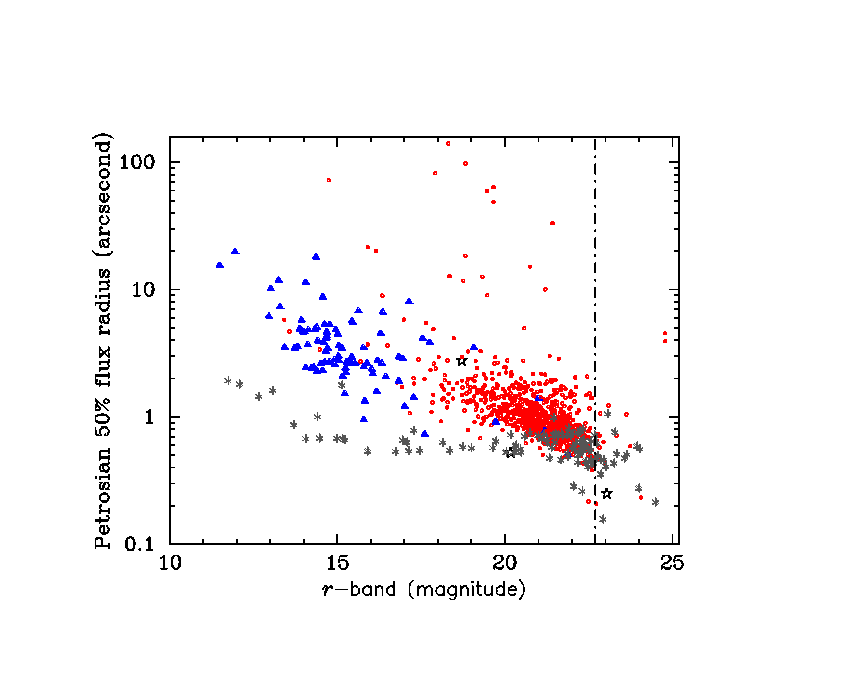}}
 \caption{The distribution of 50\% Petrosian radius as a function of extinction corrected $r$-band magnitude for the optical counterparts of our UVIT sources. The sources
 are separated into stars {\it (grey asteriks)}, galaxies {\it(red points)} and quasars {\it(black star)} by using the SDSS classification. The Coma cluster galaxies are highlighted as 
 {\it blue triangles}. The {\it dot-dashed line} represents the completeness limit for the SDSS photometric data.}
 \label{rmag-radius}
 \end{figure}

 The SDSS photometric catalogue is complete to $r = 22.7$ mag\footnote{https://www.sdss.org/dr14/imaging/other\_info/}. The optical counterparts for 968/1308 UVIT sources 
 were found in the Sloan Digital Sky Survey (SDSS, data release 14) database, of which only 923 ($\sim 70$ per cent) fall within the 
 SDSS completeness limit. We considered a maximum radius of $5^{\prime\prime}$ to find an optical counterpart for each UVIT source in our catalogue. Of these, 
 96.3 per cent of the optical counterparts were found within $3^{\prime\prime}$ of the UVIT source position. 
 
 Amongst others, the SDSS photometric pipeline provides object type for most of the matched sources brighter than the
 magnitude completeness limit. Combining information from the SDSS and the NASA Extragalactic Database 
 (NED) \footnote{https://ned.ipac.caltech.edu/}, we could identify 969/1308 objects: 114 stars, 852 galaxies and 3 quasi-stellar objects (QSO), respectively. 
 However, only 957 of these sources meet the magnitude completeness limit for SDSS. We also note that the SDSS object type for some sources was modified based on 
 visual inspection, spectroscopic information or additional data obtained from NED. Furthermore, spectroscopic redshifts for 95 sources were found in the SDSS database, 
 while an additional 39 were obtained from NED. In Table~\ref{cat} we show a sample of few rows from the full catalogue of 1308 UVIT sources detected in this image, the 
 complete version of which is available online.
 
  \begin{table*}
 \caption{The object ID, sky coordinates, UVIT flux and the uncertainty in the flux, object type (1: QSO; 3: galaxy; 6: star), $r$-band magnitude, plate, MJD and Fiber ID from SDSS, 
 redshift and the source of redshift (0: Not available; 1: SDSS; 2: NED) for all the 1308 sources detected in our image. (A complete version of this table is available online)}
 \begin{center}
 \begin{tabular}{ rccccccccccc }     
 \hline
   ID  & $\alpha$ & $\delta$  & Flux  ($10^{-15}$ &  $\Delta$(Flux) ($10^{-15}$    & Object  &  $m_r$  & Plate & MJD  & Fiber ID  & $z$  & Source of  \\ 
         &    (J2000)           &   (J2000)      &  ergs/s/cm$^2$/\AA)  & ergs/s/cm$^2$/\AA)  & type      &  mag&  &   &      &   & redshift ($z$) \\ \hline
1065  &  195.063  &  27.893    &  0.1150   &  0.0046  &  3  &   19.00   & -   &   -   & -   &  0.281  &  2\\
 238  &  194.819  &  28.132  &    0.1976   &  0.0062  &  3   &  18.75  &  -   &   -  &  -  &  -  &  0  \\
 790  &  195.159   & 27.989   &   0.1197  &   0.0047  &  3  &   18.36  &  -  &    -  &  -  &   0.168  &  2\\
  72  &  195.015   & 28.219  &    0.1502  &   0.0054  &  3  &   19.11  &  -   &   -  &  -  &  -  &  0\\
 150  &  195.019  &  28.184   &   0.1720  &   0.0058   & 3  &   18.92  &  -   &   -  &  -  &  -  &  0\\
 884  &  194.899  &  27.959  &    0.1820   &  0.0059  &  3   &  11.98  &  2240   &  53823  &  585  &   0.024   & 1\\
 307   & 194.844   & 28.129   &   0.1907  &   0.0061  &  3   &  17.90   & 2240  &   53823   & 576   &  0.194  &  1\\
1195  &  195.091   & 27.845  &    0.1249  &   0.0048  &  3  &   18.91  &  -  &    -   & -  &  -  &  0\\
  49  &  195.106   & 28.229   &   0.2224   &  0.0064  &  3   &  18.23  &  -   &   -  &  - &   -  &  0\\
   1  &  195.026   & 28.253   &   1.6038   &  0.0179  &  3   &  17.18  &  2241 &    54169  &  407  &   0.021   & 1\\ \hline
 \end{tabular}
 \end{center}
 \label{cat}  
 \end{table*}
  
 In Fig.~\ref{rmag-radius}, the Petrosian radius containing 50 per cent of the total flux in the $r$-band ($R_{50}$) is shown as a function of the extinction corrected $r$-band 
 magnitude for all the optical counterparts, sub-classified into the Coma cluster galaxies, other galaxies, quasars and stars, respectively. 
 The Coma cluster members have been identified assuming $z_{Coma} = 0.023$ and velocity dispersion ($\sigma_v$) of 1000 km s$^{-1}$. 
 These criteria yield 79 member galaxies within $\pm 3\sigma_v$ 
 as members of the Coma cluster. All the other extended sources, with or without redshift, are addressed as `other galaxies' henceforth, although it 
 is possible that some of the fainter galaxies in this category might be members of the Coma cluster. 

   Fig.~\ref{rmag-radius} shows that most of the bright galaxies found in these UVIT data are members of the Coma cluster. These also happen 
 to be the largest galaxies in our sample. Most of the stars brighter than $r \sim 20$ mag can be separated out from the galaxies in this plane. 
 While one of the quasars coincides with stars, the second follows the galaxies, and the third one lies in the region occupied by the faint stars in this $r$-band magnitude-radius plane. 

\begin{figure}
 \includegraphics[scale=0.16]{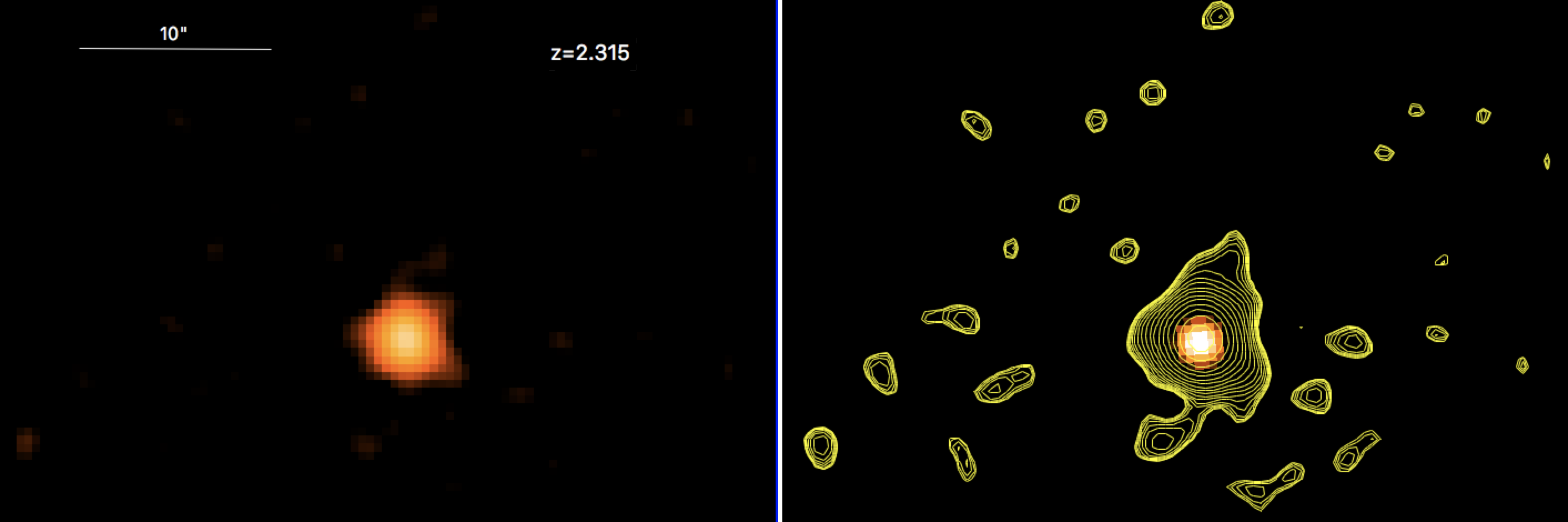}
 \caption{The UVIT {\it FUV (left)} and {\it FUV} contours overlaid on the optical $r$-band image {\it (right)} of the quasar B1257$+$280 ($z=2.315$). }
 \label{qso}
 \end{figure}

 It is also worth mentioning here that the three QSOs have $z = 0.383, 0.513$ and $2.315$, respectively. While the last one has spectroscopic redshift available in the SDSS 
 database, redshift for the other two were obtained from NED. The quasar B1257$+$280 ($z=2.315$) shown in Fig.~\ref{qso} deserves a special mention, since this might be the highest redshift 
 object observed by the UVIT to date. The multi-band optical image shows a blue star-like object, while the faint {\it FUV} emission at the edges show elongation in flux contours in the north-south direction. 
 This QSO is known to be a Lyman $\alpha$ absorber \citep{garnett17}.
 
 \section{Properties of the detected FUV sources}
 \label{props}
 
  \begin{figure}
  \adjustbox{trim={.085\width} {.001\height} {0.001\width} {.001\height},clip}
 {\includegraphics[scale=0.45]{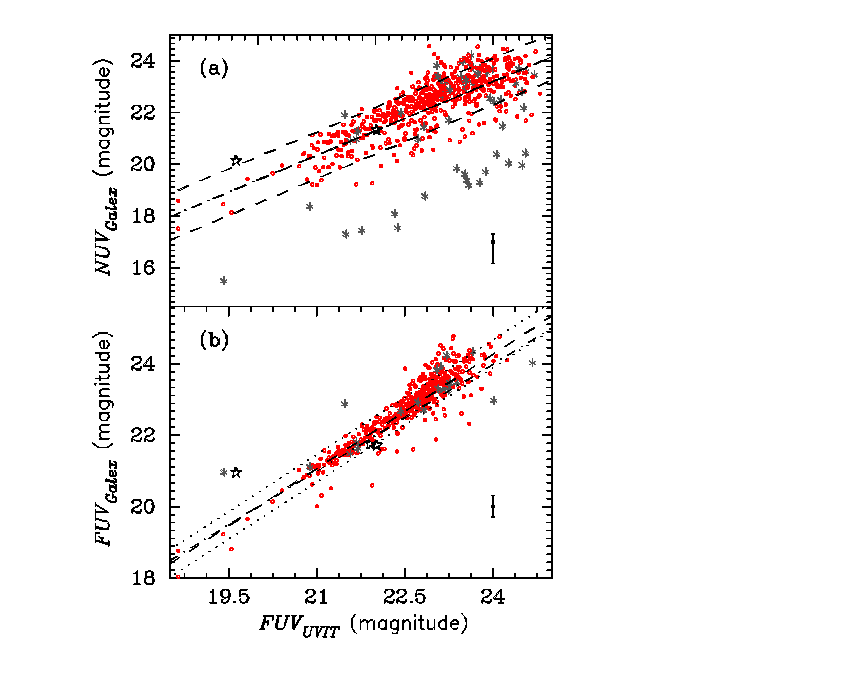}}
 \caption{The distribution of magnitude of \g NUV sources and (b) \g FUV sources plotted as a function of the FUV magnitude of the corresponding UVIT source, respectively. 
 All magnitudes are corrected for extinction due to Milkyway. The {\it red points} denote galaxies, {\it grey asterisks} represent stars and QSOs are shown as {\it stars}, respectively. 
 The {\it dot-dashed line} in both panels represents a perfect correlation, while the {\it dashed lines} represent the least square fit to the data along with $1\sigma$ deviation in it. 
 Typical uncertainty along each axis is shown in the bottom right corner of the panels. This figure therefore shows that the FUV fluxes of the matched 
 sources are consistent between the UVIT BaF$_2$ band and both the wavebands of the \g mission. A marginal discrepancy of increasing magnitude in \g FUV, can be noticed towards the 
 faint end, which is however well within $\sim 3\sigma$ for most of the sources.}
 \label{uvit-galex}
 \end{figure}
 
 The Coma field studied here has previously been observed with the {\it Galex} mission. So as a sanity check we compared our catalogue to the one created from {\it Galex} images. 
 In order to do so, we made use of the processed images downloaded from the galexview\footnote{https://galex.stsci.edu/GalexView/} page. The comparison images have an 
 exposure time of 25656.55 seconds in $NUV$ and 18620.45 seconds in the $FUV$ band, respectively (Object ID: GI5\_025001\_COMA\_0002). {\sc sextractor} was used
 to identify the sources in the {\it Galex} images. We then searched for a {\it Galex} sources within $3^{\prime\prime}$ of each of the UVIT sources independently in the near and far UV catalogues. This resulted in 464 
 FUV and 599 NUV sources matched between the UVIT and {\it Galex} catalogues. 
    
 The distributions of magnitudes of all the matched sources in the two \g wavebands are shown in Fig.~\ref{uvit-galex}. Both stars and galaxies as identified in the previous section
 have well correlated FUV fluxes in the UVIT and the \g bands, albeit with the expected scatter which is a consequence of mismatched wavelength range covered by the three 
 wavebands analysed here. We note that in the $FUV_{UVIT}$--$NUV_{Galex}$ plane, bright `stars' are well separated from the galaxies. The best-fit least square relation between the UVIT and \g bands for these data as shown in Fig.~\ref{uvit-galex} are:
 \begin{equation}
 \begin{split}
 Galex_{NUV} = 0.3808 +  0.9514UVIT_{FUV}~~\pm 0.8995, \\
 Galex_{FUV} = -1.3338 +  1.0672UVIT_{FUV}~~\pm  0.3851
 \end{split}
 \end{equation}
 
 In Fig.~\ref{uvit-galex} we show that despite good general agreement between the UVIT $FUV$ and \g bands, there are subtle discrepancies, particularly at the faint end. 
 We note that  
 the scatter is higher at the faint end in both the distributions. In fact, the standard deviation in the $FUV_{UVIT}$--$NUV_{Galex}$ correlation seems to be dictated by the faint end
 scatter. The correlation between $FUV_{UVIT}$ and $FUV_{Galex}$ is relatively better, but we observe a slight upturn in the distribution for galaxies fainter than 
 $FUV_{UVIT} \sim 22.6$ mag. The trend implies that these faint galaxies detected by both UVIT and {\it Galex}, are fainter in \g relative to UVIT. This might be a consequence of the poor 
 resolution of \g relative to UVIT, but further investigation may be required if similar trends are observed in other UVIT fields as well. It should be noted, however, that the typical uncertainty 
 in the \g~bands is $\sim 0.44$ and $0.57$ magnitude in the $FUV$ and $NUV$ bands, respectively, which is an order of magnitude higher than the uncertainty of $\sim 0.03$
 magnitude estimated for the UVIT BaF$_2$ band. 
  
  \begin{figure}
  \adjustbox{trim={.085\width} {.001\height} {0.001\width} {.001\height},clip}
 {\includegraphics[scale=0.32]{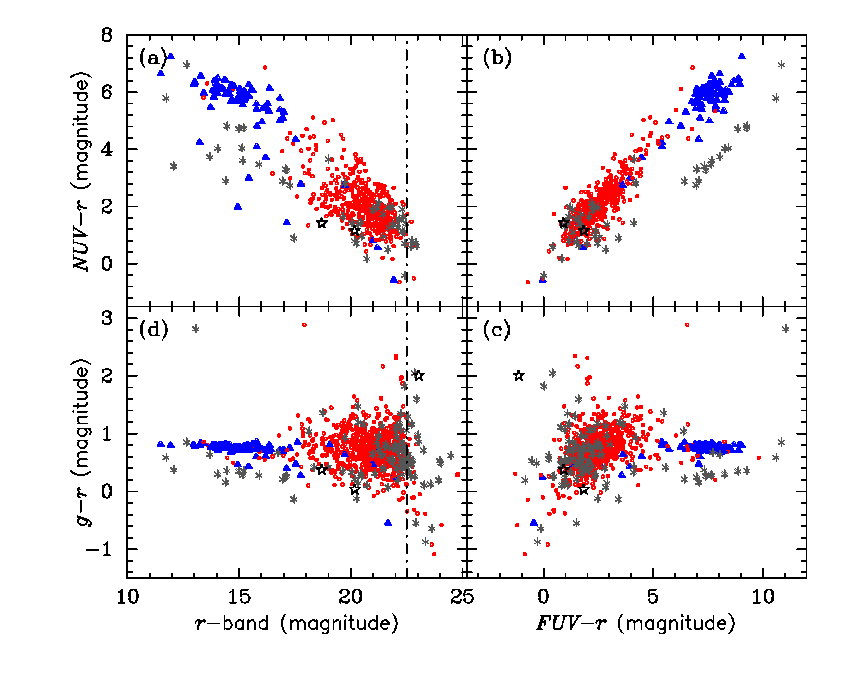}}
 \caption{The colour-colour and colour--magnitude distributions for all UVIT sources along with their optical and \g counterparts, respectively in the (a) $NUV-r$ vs $r$-band magnitude, 
 (b) $NUV-r$ vs $FUV-r$, (c) $FUV-r$ vs $g-r$, and (d) $g-r$ vs $r$-band magnitude plane. The symbols are same as in Fig.~\ref{rmag-radius}. The $FUV$ magnitudes are 
 from UVIT, while the NUV magnitudes are from the \g mission. Given the heterogeneous nature of the data plotted here, the number of data points plotted in each quadrant vary 
 as mentioned in Sec.~\ref{data}. The Coma cluster galaxies are distributed coherently in the brighter, redder region of the respective planes, while other galaxies are relatively fainter
 and bluer in comparison, with more scatter. It is notable that the other galaxies and stars are not easily distinguishable in all of these planes, particularly at the faint end with one 
 exception. A handful of stars redder than $NUV-r \sim 2$ mag and $FUV-r \sim 5$ mag, are somewhat separated from galaxies (panels (a), (b) and (c)). } 
 \label{col-mag}
 \end{figure}

 In Fig.~\ref{col-mag} we show the distribution of the UVIT sources in various optical, UV and composite colour-magnitude and colour-colour planes. 
 Note that information in all wavebands is not available for all the sources, hence different number of sources are plotted 
 in each of the four panels. The Coma cluster galaxies
 form a well defined sequence in the optical {\it g-r} vs $r$-band colour-magnitude plane (Fig.~\ref{col-mag} (d)), but the other fainter galaxies show a larger scatter.
 On the other hand, all galaxies follow a continuous sequence in the plane mapped by the {\it NUV-r} colour and $r$-band magnitude (Fig.~\ref{col-mag} (a)). 
 It is interesting to note that most of the stars brighter than $r \sim 19$ mag span similar range of r-band magnitudes as the bright Coma cluster galaxies, yet have relatively bluer 
 {\it g-r}, {\it NUV-r} and {\it FUV-r} colours.   
 
 The Coma cluster galaxies also follow a well defined sequence in the {\it g-r} vs {\it FUV-r} colour-colour plane (Fig.~\ref{col-mag} (c)), and the {\it NUV-r--FUV-r}
 plane (Fig.~\ref{col-mag} (b)), respectively. All other galaxies, although fainter than their Coma cluster counterparts, seem to follow the same correlation in the {\it NUV-r} vs 
 {\it FUV-r} plane. 
 Just like in panels (a) and (d), in the plane mapped by the {\it NUV-r} and {\it FUV-r} colours, a small subset of bright stars seems to follow a correlation just like the 
 galaxies but have {\it FUV-r} colours redder by $\sim 2$ magnitude at fixed {\it NUV-r} colour, relative to the galaxies. As mentioned earlier, many of the objects at the faint 
 end classified as stars by the SDSS photometric pipeline may be high-z quasars. Also, many of the `other' galaxies falling in the same region of the colour-colour and colour-magnitude 
 planes as the Coma cluster galaxies may also be member galaxies of the Coma cluster. 
 
  \begin{figure}
  \adjustbox{trim={.085\width} {.001\height} {0.001\width} {.4\height},clip}
 {\includegraphics[scale=0.33]{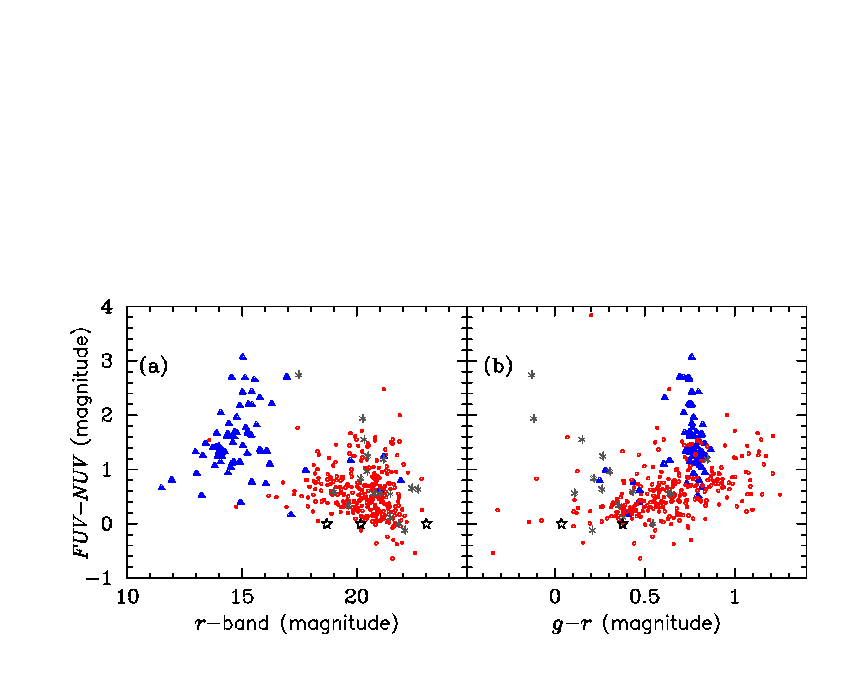}}
 \caption{The distribution of all the UVIT sources having optical and \g counterparts in the (a) {\it FUV-NUV} vs $r$-band and (b) {\it FUV-NUV} vs {\it g-r} colour-colour plane, 
 respectively. The symbols are same as in Fig.~\ref{rmag-radius}. Most of the Coma cluster galaxies fall in well defined regions of the space
 dictated by their optical properties, while stars and quasars co-exist with the other, relatively fainter galaxies.} 
 \label{fnuv}
 \end{figure} 
 
In Fig.~\ref{fnuv} we show the distribution of various sources in the plane mapped by the UV colour and the $r$-band magnitude and the {\it g-r} colour, respectively. The 
difference in the $r$-band luminosity of the Coma cluster galaxies and other galaxies leads to a clear separation between the two sub-samples of galaxies in the 
$r$ vs $FUV-NUV$ colour-magnitude plane. On the other hand, the Coma cluster galaxies form a tight sequence in the $g-r$ vs $FUV-NUV$ colour-colour plane 
 (Fig.~\ref{fnuv} (b)), an expected signature from the optical colour-magnitude diagram of clusters. We note however, that stars and quasars remain indistinguishable from 
 galaxies in these planes. 
 
 To summarise, this analysis shows that (i) the {\it FUV} flux from the UVIT BaF$_2$ filter data are well correlated with the \g bands, (ii) Coma cluster galaxies are much brighter,
 and hence well separated from the other galaxies in the image in almost all the colour-colour and colour-magnitude planes explored here, and (iii) stars and quasars
 have identical photometric properties as the other, relatively fainter galaxies.

  
 \section{FUV sources with unusual morphology}
 \label{distorted}
 
 We found many galaxies with unconventional morphology in the UVIT $FUV$ image by visualizing the UVIT image simultaneously with the optical $r$-band image. 
 Objects showing tails, rings or other features suggesting the impact of environment-related mechanism, and galaxies which showed morphology very different from their optical
 counterparts, were shortlisted and analysed further. The galaxies for which subsequent investigation showed that the `features' were background objects, were then 
 dropped, finally leaving a set of {\bf sixteen bright and seven faint} galaxies with distorted morphology. Some of these 
 galaxies have been subjects of other optical and ultraviolet studies 
 in the literature \citep[e.g.][]{yagi10,smith10}, with most of them showing an `unusual' morphology in the UV bands only. We describe each of them briefly below, incorporating 
 relevant information from the literature where available. All the objects are listed in Table~\ref{tab:HI}, and their UVIT {\it FUV} and optical $r$-band images are shown in
 Figs.~\ref{images}-\ref{images6}. The GMP identification in the first column of Table~\ref{tab:HI} is from \citet{gmp83}. 
 Redshift information is available only for the first sixteen bright galaxies with unusual {\it FUV} morphology in the Table.
 Where available, Table~\ref{tab:HI} also includes the stellar masses of galaxies from \citet{mendel14} or the GSWLC, and HI-mass fraction estimated 
 by applying artificial neural network on SDSS optical data \citep{teimoorinia17}.

  \begin{table*}
 \caption{The common name, sky coordinates and redshift for the objects with distorted morphology found in the FUV image. 
 Where available, the last two columns show the stellar mass and the HI-to-stellar mass ratio of the objects. The last seven objects 
 are optically-faint galaxies without any redshift information.}
 \begin{center}
 \begin{tabular}{ lcccccc }     
 \hline
  Common   &  Right Ascension & Declination &  Redshift     & log $M*$             & log $M_{HI}/M*$  \\ 
  name          &  (J2000)      &  (J2000)   &                     &  ($M_\odot$)  &               \\  \hline 
 GMP 3016   & 195.005 & 28.082 & 0.0259 & - & - \\ 
 PGC 044679  & 194.979 & 28.128 & 0.0252 & 10.253 & 0.0878 \\ 
 GMP 2787  & 195.076 & 28.059 & 0.0332 &   9.390 & -0.2226 \\ 
 GMP 3523    & 194.843 & 28.129 & 0.1943 & - & - \\ 
 GMP 2910  & 195.038 & 27.866 & 0.0177 & 9.568 & 0.1665 \\ 
 NGC 4867    & 194.813 & 27.971  & 0.0161 & 10.17 & -0.6905 \\  
 NGC 4895   & 195.075 & 28.202 & 0.0283  & - & - \\ 
 GMP 2584  & 195.148 & 28.146 & 0.0182 &   9.972 & -0.3045 \\   
 GMP 2559  & 195.158 & 28.058 & 0.0255 & - & - \\ 
 GMP 2347 & 195.247 & 27.899 & 0.0230 & - & - \\
 NGC 4907 & 195.204 & 28.159 & 0.0194 & 10.766 & - \\
 GMP 2943 & 195.026 & 28.253 & 0.0211 & 8.530 & - \\
 GMP 2989 & 195.012 & 28.241 & 0.0257 & 9.900 & -0.0648 \\
 GMP 3317 & 194.901 & 28.042 & 0.2747 & - & - \\
 GMP 3170 & 194.945 & 27.974 & 0.0314 & 10.674 & -0.9705 \\ 
 GMP 2956 & 195.023 & 27.807 & 0.0219 & 10.255 & -0.3113 \\
 \hline
GMP 2717 & 195.095 & 27.821 & - & - & - \\
Subaru-UDG & 195.064 & 27.836 & - & - & - \\
-         & 194.953 & 27.797 & - & - & - \\
GMP 2476 & 195.190 & 27.918  & - & - & - \\
GMP 2454 & 195.199 & 27.994 & - & - & - \\
GMP 2320 & 195.259 & 28.102 & - & - & - \\
-        & 195.205 & 28.064 & - & - & - \\ \hline
 \end{tabular}
 \end{center}
 \label{tab:HI}  
 \end{table*}

 \subsection{GMP 3016}
\begin{figure*}
  \includegraphics[scale=0.341]{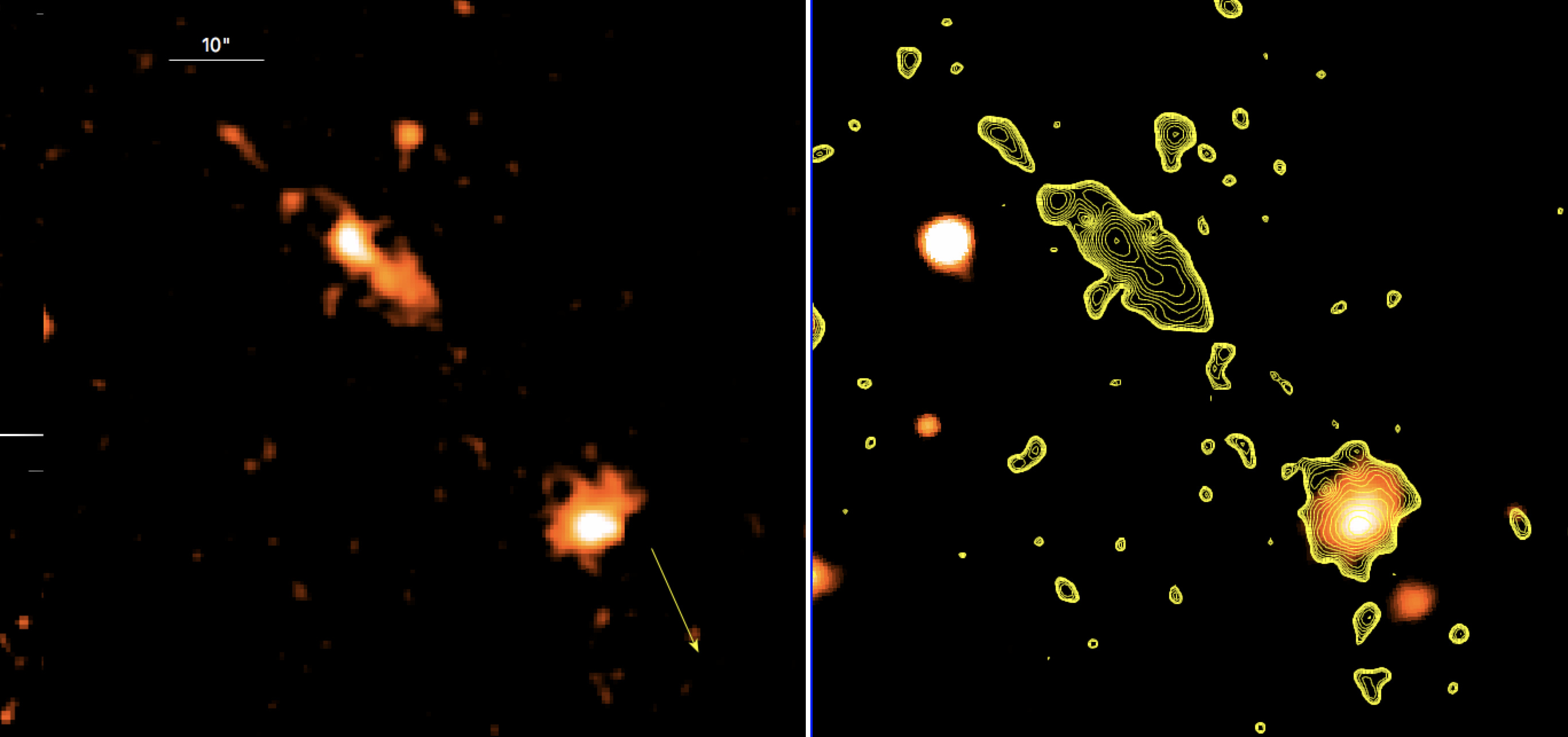}
  \includegraphics[scale=0.35]{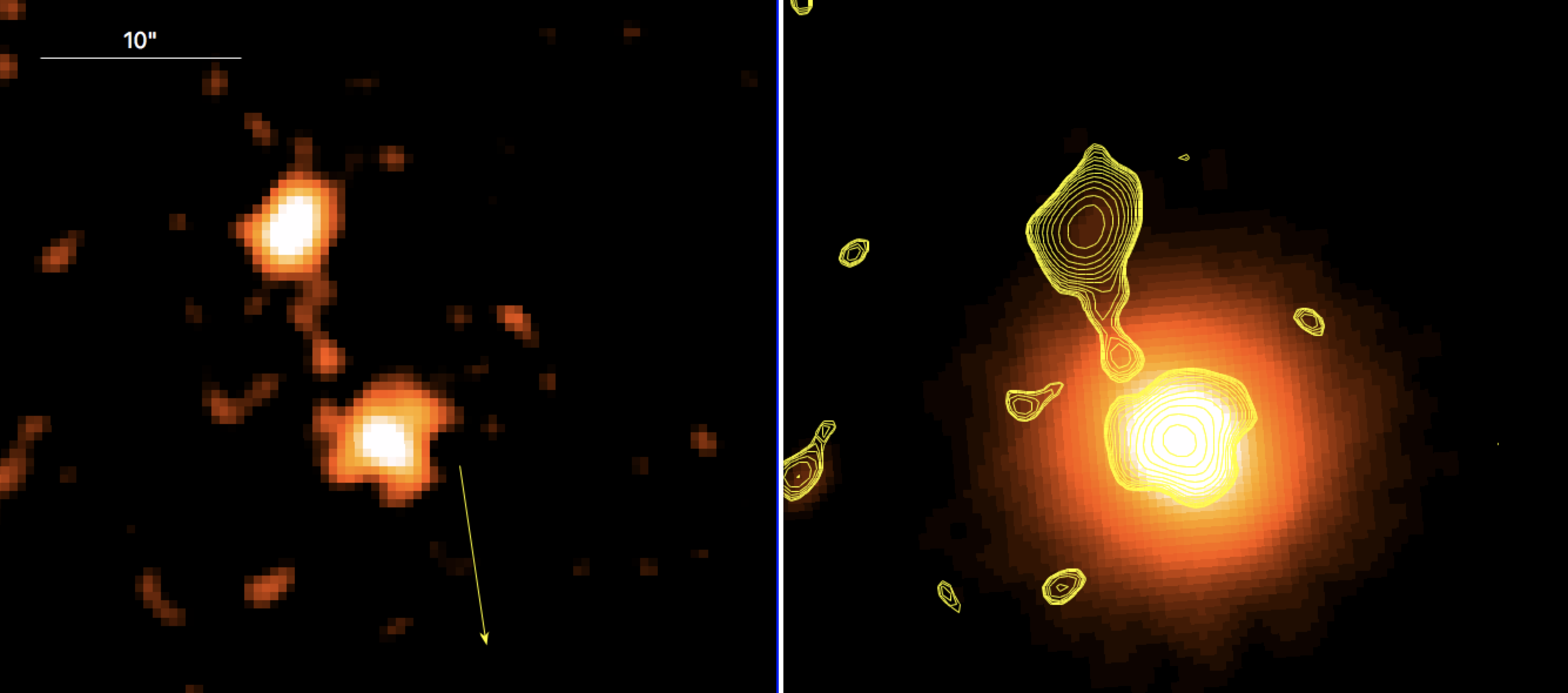}
  \includegraphics[scale=0.352]{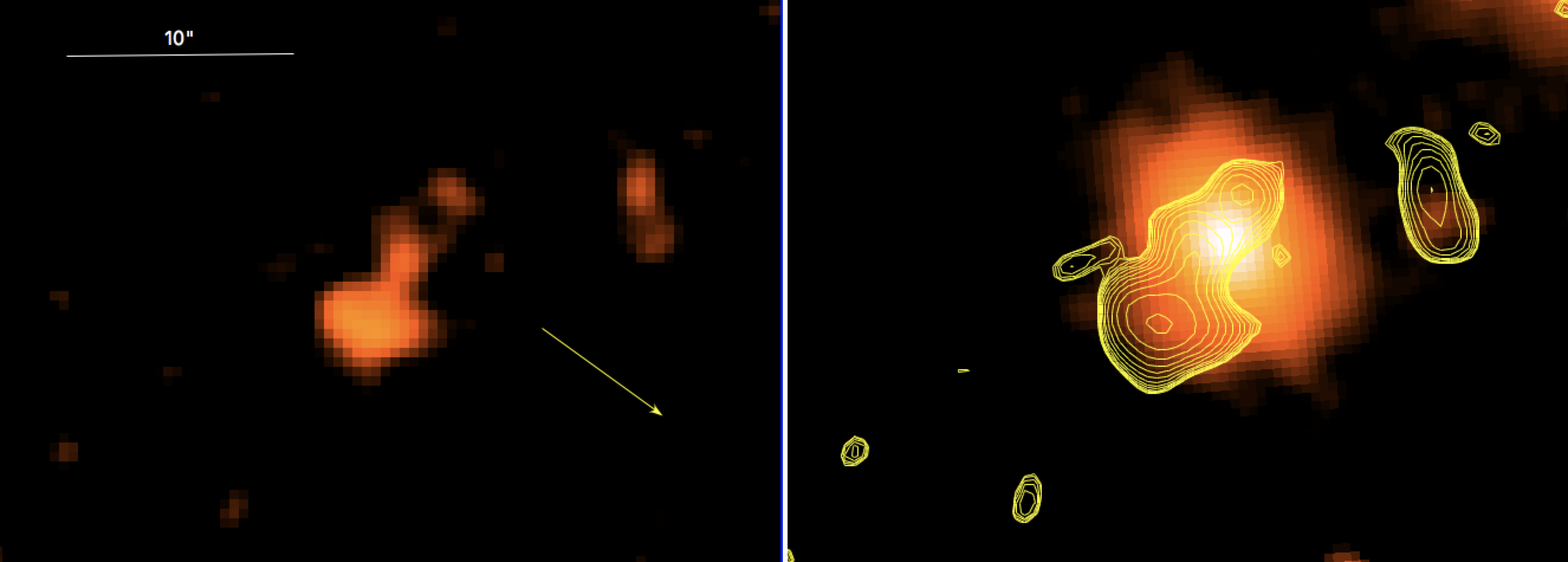}
 \caption{This figure shows {\it (left):} the UVIT FUV image, and {\it (right):} the $r$-band optical image with FUV contours overlaid for (a) GMP 3016, (b) PGC 044679, and 
 (c) GMP 2787, respectively from top to bottom. The contours are smoothed over 4 pixels and created such that the outermost contour is 2-3$\sigma$ above the typical background of the UVIT image.
 The arrow points towards the cluster centre. Images are oriented such that north is up and east is on the left of the image. 
 These images evidently show the difference in morphology of the young, UV emitting stars and the generic distribution of other stellar populations in these galaxies. 
 The impact of the cluster environment can also be noticed for galaxies which show UV emission away from the plane of the galaxy, often with no optical counterpart. }
 \label{images}
 \end{figure*}

 This is an extremely blue, faint irregular galaxy as observed in the optical wavebands. \citet{yagi10} have reported `fireball' features i.e., extended \ha~emission 
 following knots of blue stars \citep{yoshida08} from GMP 3016 (see the right panel of Fig.~\ref{images}(a)). On the other hand, \citet{smith10} found this to be a `less convincing case' of a stripping galaxy in the combined {\it NUV$+$FUV} image obtained with {\it Galex}. 
 Our UVIT image (Fig.\ref{images}(a)) however, shows a linear structure in the north-eastern part of the {\it FUV} image clearly extending towards GMP 3016. In {\it FUV}, this trail is 
 detached from the galaxy and extends upto $\sim 26$ kpc in the north-east direction\footnote{All galaxies shown in Figs.~\ref{images}, \ref{images2} and \ref{images3} are oriented such that north is up and east is left of the image.},
 with the cluster centre $< 200 h^{-1}$ kpc away parallel to the direction of the alignment of knots. In our opinion, the morphology of the narrow structure extending in the 
 north-east suggests that it is likely a part of GMP 3016, but without any optical counterpart. 
  
 \subsection{PGC 044679}
 This is a barred S0 galaxy member of the Coma cluster \citep{lansbury14}. The FUV image (Fig.~\ref{images}(b)) shows two discrete components, one centred at the galaxy's nucleus and the 
 other in the north-east. The nuclear component (C1 in Fig.~\ref{images}(b)) has an interesting morphology pointing in orthogonal directions, one of which points in the direction of the cluster centre
 which is at a distance of $\sim 256 h^{-1}$ kpc from the galaxy. The second component C2, however, remains to be 
 further investigated to confirm whether it also corresponds to PGC 044679, is a satellite galaxy or a random background object. Unlike GMP 3016, this is therefore one of 
 our less convincing cases of multiple component UV emission. Table~\ref{tab:HI} shows that this galaxy is detected in HI (Table~\ref{tab:HI}).
  
 \subsection{GMP 2787}
 This is an irregular galaxy ($z=0.033; 9974$ km s$^{-1}$), having radial velocity slightly above the maximum threshold considered here for selecting the members of the 
 Coma cluster. It has a stellar mass of $9.39 M_\odot$ \citep{teimoorinia17}. This galaxy appears to be very faint in the {\it Galex} images. The UVIT image (Fig.~\ref{images}(c)) of this galaxy 
 shows an irregular morphology, very different from the optical extent of the galaxy, and pointing in a direction away from the cluster centre which is $\sim 236 h^{-1}$ kpc away. 
  
  \subsection{GMP 3523}
 This is a star-forming galaxy at $z=0.194$ with tail like features observed on the top and south-western edge of the galaxy (Fig.~\ref{images2}(a)). All the tail like features are invisible 
 in the optical image, and may be tidal tail(s) of this galaxy, although this needs to be confirmed in further investigation. The south-western extension was also seen in the {\it Galex} image 
 \citep[e.g.][]{donas95}, but appeared as a fainter feature relative to the UVIT image. 
     
 \subsection{GMP 2910}
 This is the most well studied of our sources with unusual FUV morphology. Several authors have reported extended emission associated with this post-starburst galaxy commonly known as D100
  \citep[$z=0.0177$; ][]{yagi07,smith10,yagi10,cramer19}. This source was also reported as KUG $1257+281$ in the Kiso Survey for UV-excess galaxies \citep{takase93}. The FUV emission observed 
 in the UVIT image (Fig.~\ref{images2}(b)) coincides well with the extended H$\alpha$ emission \citep{yagi07}. 

 GMP 2910 is rich in molecular \citep{jachym17} and atomic gas \citep{teimoorinia17}, with a considerable HI-to-stellar mass fraction ($> 0.16$; Table~\ref{tab:HI}). The
 extended {\it FUV} emission points in the northeast direction almost perpendicular to the vector pointing in the direction of the cluster centre. The spectacular narrow tail observed with the Subaru telescope's 
 Suprime-Cam in the H$\alpha$ narrow band filter \citep[see Fig. 2 of][]{cramer19} is well aligned with the {\it FUV} feature observed with the UVIT. 

 This galaxy is only $\sim 240 h^{-1}$ kpc away from the cluster centre, and hence likely a classic case of ram-pressure stripping. However, the narrow, linear structure 
 emerging from the galaxy is difficult to explain via ram pressure stripping alone. The other possibility is that the structure is formed by a dwarf galaxy or gas cloud which 
 was disrupted due to ram-pressure stripping by the ICM, or tidal forces of GMP 2910 \citep{yagi07}.  In a recent study, \citet{peluso22} explored the connection between active 
 galactic nucleus (AGN) and ram pressure stripping, in a sample of 115 ram-pressure stripped galaxies \citep[also see,][]{george19}. 
 \citet{peluso22} find that the incidence of AGN increases among ram-pressure stripped galaxies, relative to a stellar mass matched sample of cluster galaxies, thereby suggesting that GMP 2910, 
 showing a Seyfert type 2 emission based on the emission-line diagnostic diagram \citep{bpt}, may also harbour an AGN at its centre \citep[see table 4 of][also, Mahajan et al., 2010]{peluso22}.

 \subsection{NGC 4867}
\begin{figure*}
  \includegraphics[scale=0.34]{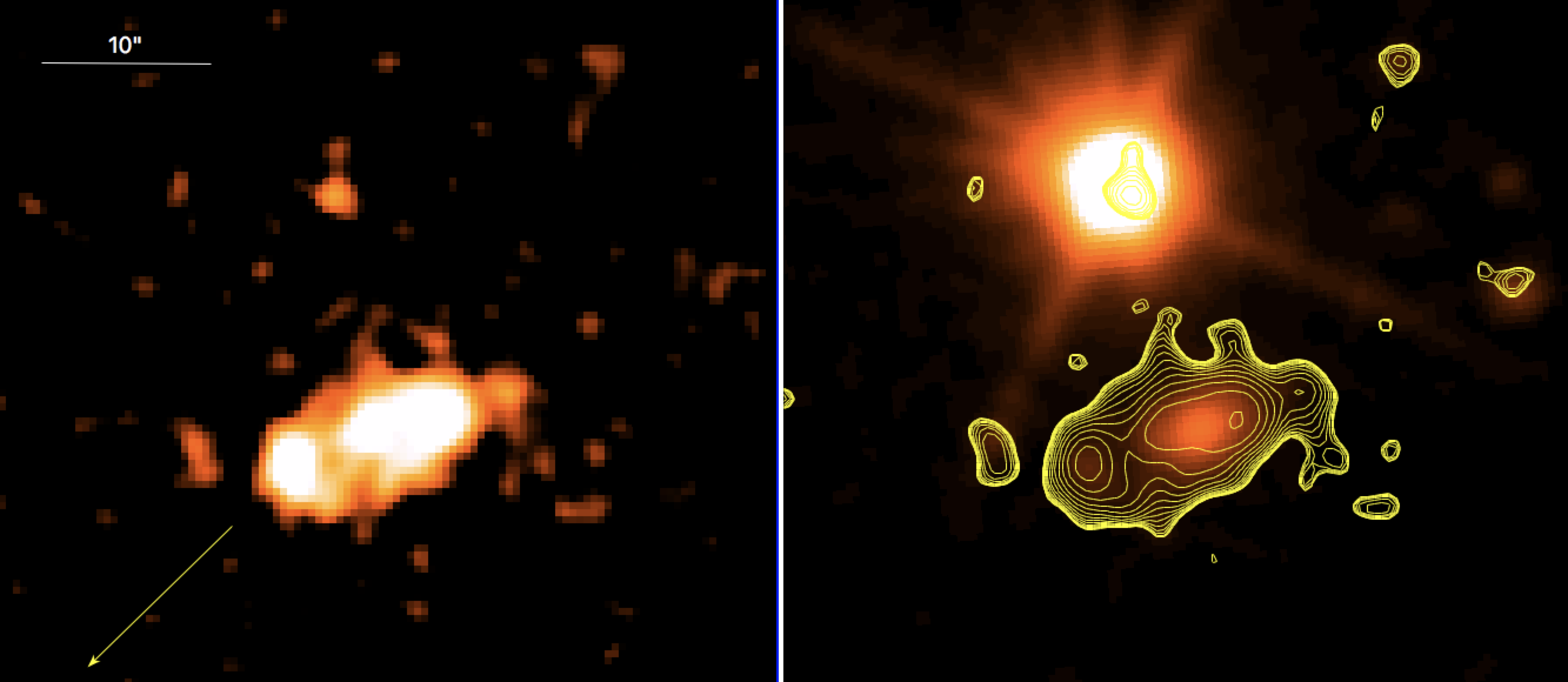}
  \includegraphics[scale=0.339]{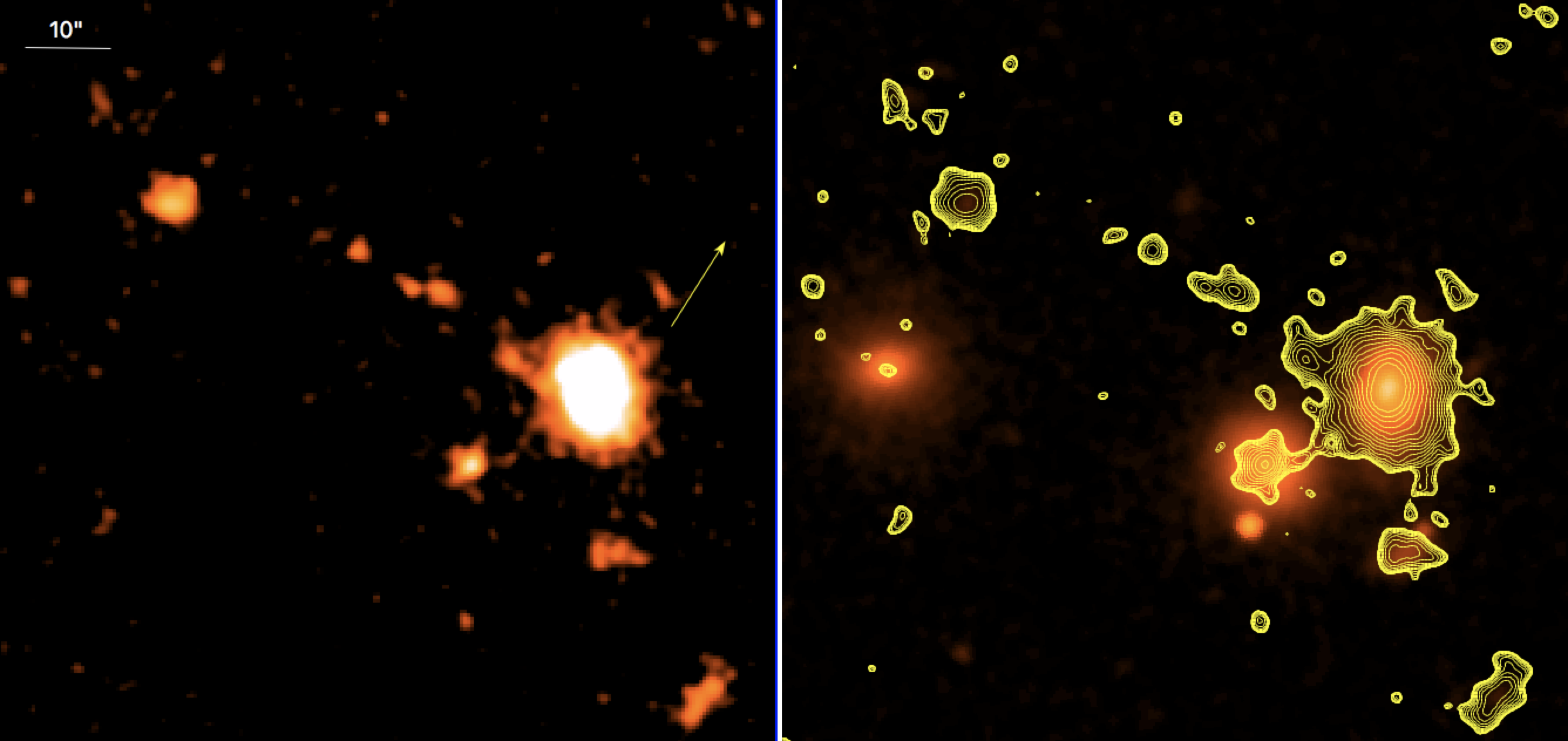}
  \includegraphics[scale=0.307]{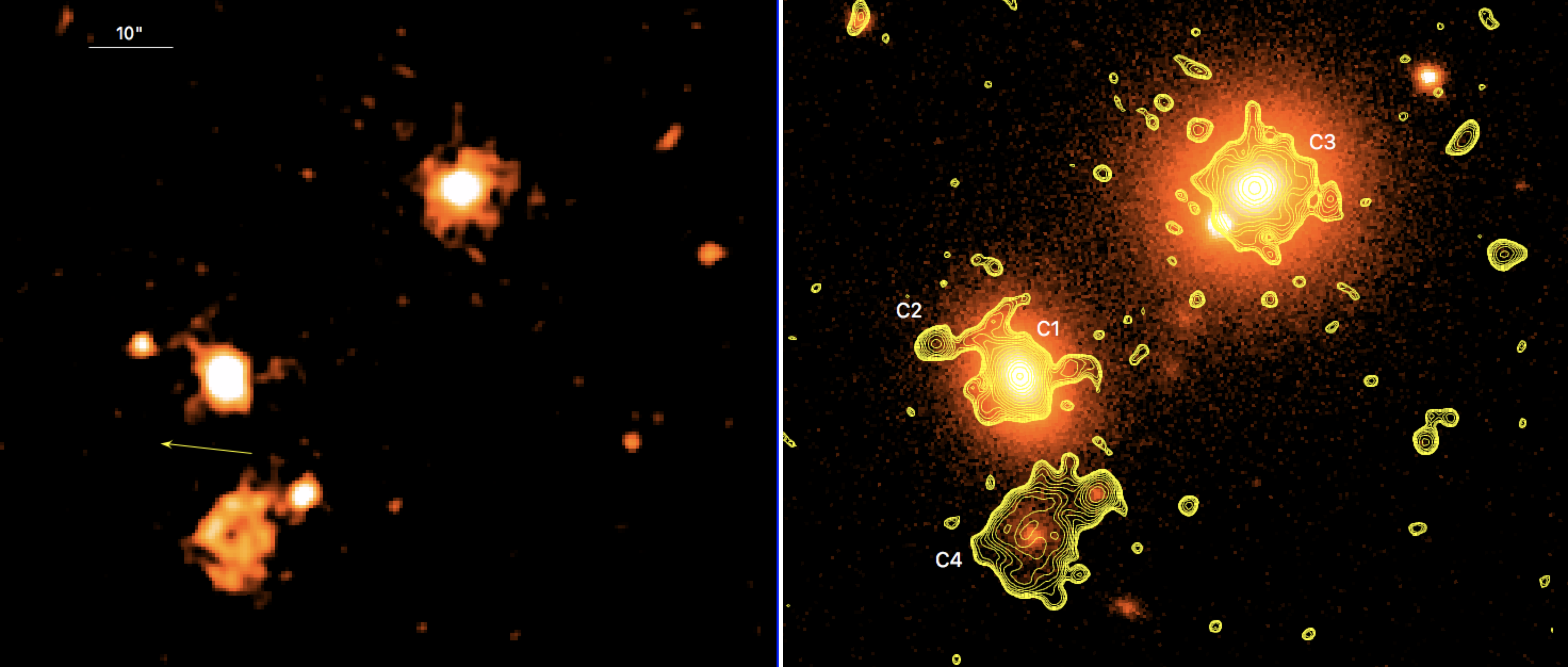}
 \caption{Same as Fig.~\ref{images}, but for (a) GMP 3523, (b) GMP 2910, and (c) NGC 4867. } 
 \label{images2}
 \end{figure*}

 This is a typical S0 galaxy in the Coma cluster ($z=0.016$), around 1.6 kpc ($5\arcsec$) from the most dominant galaxy at the centre. NGC 4867 (seen as component C1 in Fig.~\ref{images2})
 shows FUV emission extending towards west and southeast, but invisible in the optical image. A weak feature also appears to extend towards another FUV source C2 towards the left, which 
 has no optical counterpart. 
 The western tail from C1 seem to be extending towards NGC 4864 ($z=0.022$; component C3), which is also a Coma cluster member, separated in redshift space by 1800 km s$^{-1}$ from 
 NGC 4867. Hence, it is possible that the two galaxies are gravitationally influencing each other. It is notable that the cluster centre lies at a distance of $\sim 200 h^{-1}$ kpc eastward of 
 this galaxy. The third galaxy (component C4) in the image panel is a background object at $z=0.283$, but shows an interesting UV morphology.   

  \subsection{NGC 4895}
  NGC 4895 ($z=0.0283$) comprises multiple components in the FUV (Fig.~\ref{images3}(a)). While the dominant component (C1) in centred at the galaxy's nucleus, the northwestern component 
  C3 is a blue foreground star observed in the multi-band SDSS image. The relatively smaller, northeastern component C2 however, does not have an optical counterpart. The two extraplanar 
  concentrations in southwest and southeast (C4 and C5, respectively) are likely background objects, with no optical counterparts available in the SDSS database. Hence, it will be interesting 
  to explore the latter three in follow-up observations. We also note that the major axis of this galaxy is orthogonal to the direction of the cluster centre. \citet{nisbet16} found this massive
  galaxy (log$M*/M_{\odot} = 11.21$) to be hosting a low-ionization nuclear emission-line region (LINER) type AGN.

 \subsection{GMP~2584}
  This edge-on S0 galaxy in the Coma cluster, also known as PGC 044784, shows a central FUV emission reported here for the first time (Fig.~\ref{images3}(b)). The FUV emission 
  is tilted  in the direction of the cluster centre towards northwest relative to the optical disk of the galaxy. 
    
  \subsection{GMP 2559}
\begin{figure*}
   \includegraphics[scale=0.288]{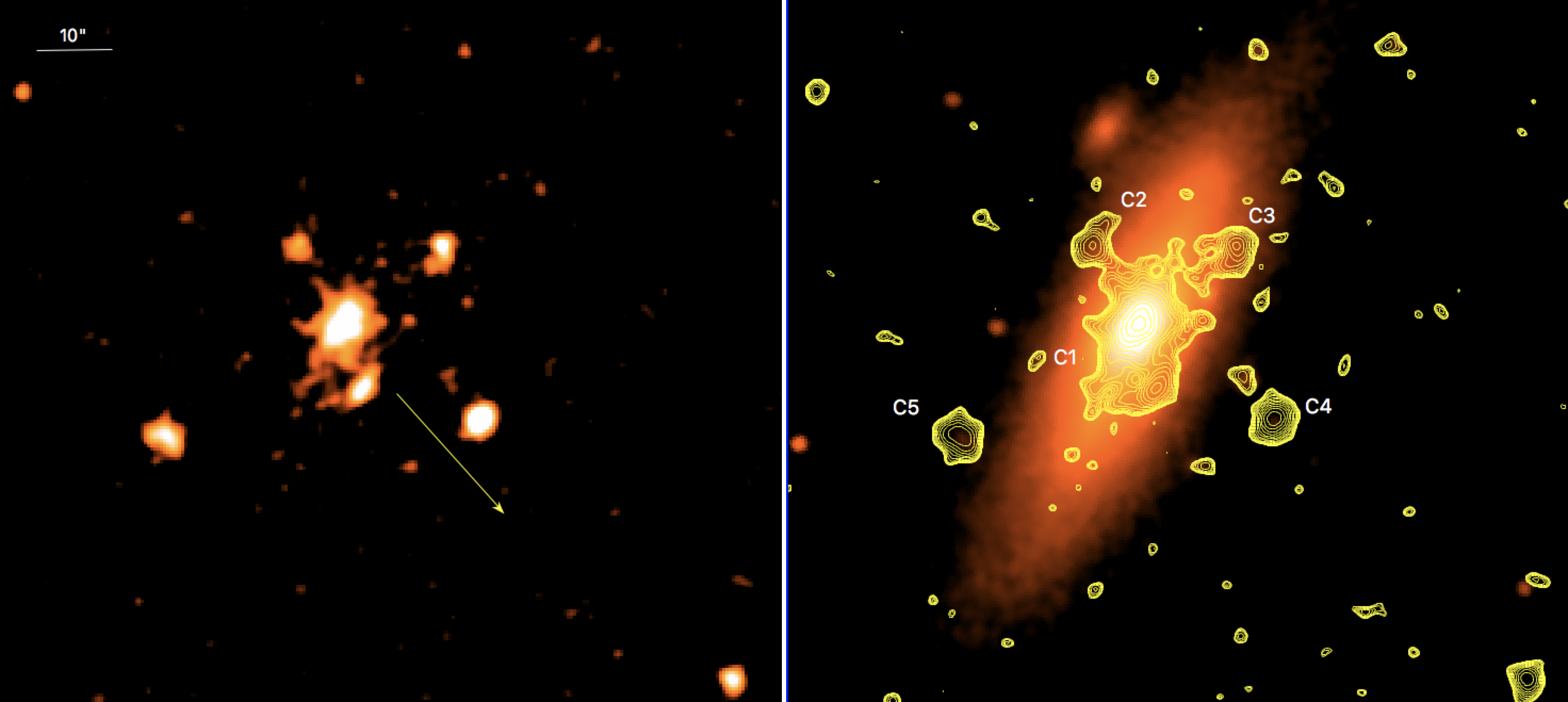}
   \includegraphics[scale=0.322]{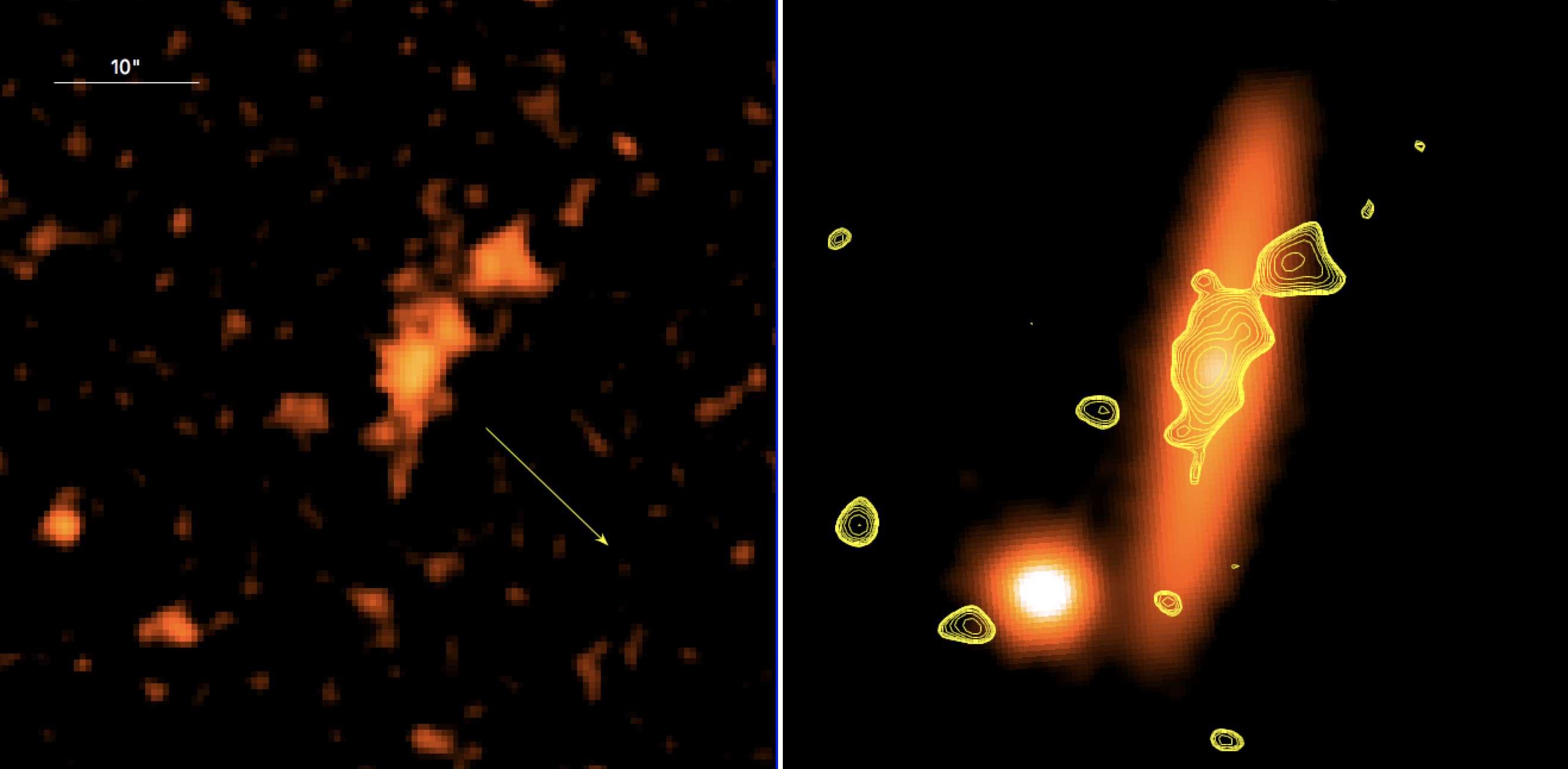}
   \includegraphics[scale=0.32]{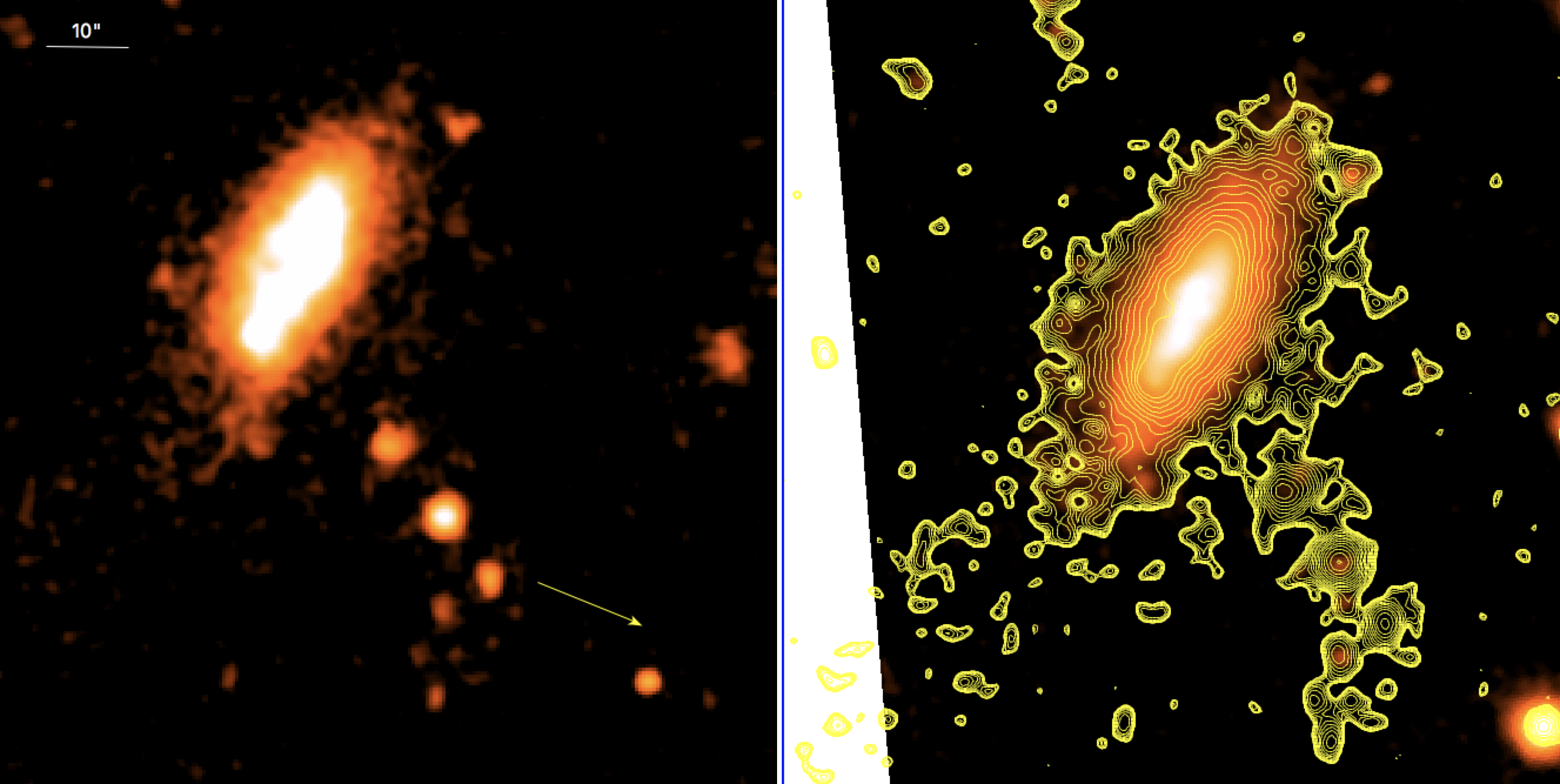}
  \caption{Same as Fig.~\ref{images}, but for (a) NGC 4895 , (b) GMP 2584, and (c) GMP 2559.   }
 \label{images3}
 \end{figure*}

  Also known as PGC 044789 and IC4040, this is a well studied blue, Sd type spiral galaxy in the Coma cluster. It is known to be an HI-deficient galaxy, having asymmetric HI distribution,
  such that most of the HI gas lies in the south-eastern part of the galaxy, while the north-western part appears depleted \citep{bravo00}. 
 The major axis of this galaxy is almost orthogonal to the cluster centre. GMP 2559 exhibits an HI tail extending towards south-east \citep{miller09, chen20}, almost coincident with the 
 extra-planar FUV emission as shown in Fig.~\ref{images3}(c). It is also noteworthy that the three ``knots" of star-forming region seen extending towards south-west in the UVIT image
 are also observed in the H$\alpha$ emission \citep[see fig.~2 of][]{chen20}, but not in the radio continuum.
  
 \subsection{GMP~2347}
 This early-type spiral galaxy in the Coma cluster shows a central {\it FUV} component (C1), along with an off-centre component C2. The latter might be a dwarf galaxy merging with GMP 2347. This galaxy is $\sim 0.45$ Mpc away from the cluster centre. Another faint galaxy GMP~2333 (component C3) can also be seen in the north-east direction relative to the galaxy in Fig.~\ref{images4}(a). In the absence of redshift information, however, it 
 is not possible to test if this faint blue galaxy is interacting with GMP 2347 or is a member of the Coma cluster.
 
 \subsection{NGC~4907}
 This face-on barred spiral galaxy in the Coma cluster, also known as GMP~2441, shows distinct {\it FUV} emission, unlike the symmetric spiral arms observed in the optical waveband. This is a LINER galaxy \citep{mahajan10, toba14}. Fig.~\ref{images4}(b) shows that unlike the optical, the 
 UVIT {\it FUV} image shows clear asymmetry between the northeast and southwestern regions of the galaxy. It is possible that gas is being pushed to the northeastern face as galaxy moves towards the cluster centre $\sim 0.44$ Mpc away. Another interesting feature is the C-shaped hole seen in the {\it FUV} emission surrounding the bar at the centre. This cavity may be a consequence of the bar suppressing star formation 
 in the centre just like in Messier 95 \citep{george19a}.
 
 \subsection{GMP~2943 and 2989}
\begin{figure*}
  \includegraphics[scale=0.329]{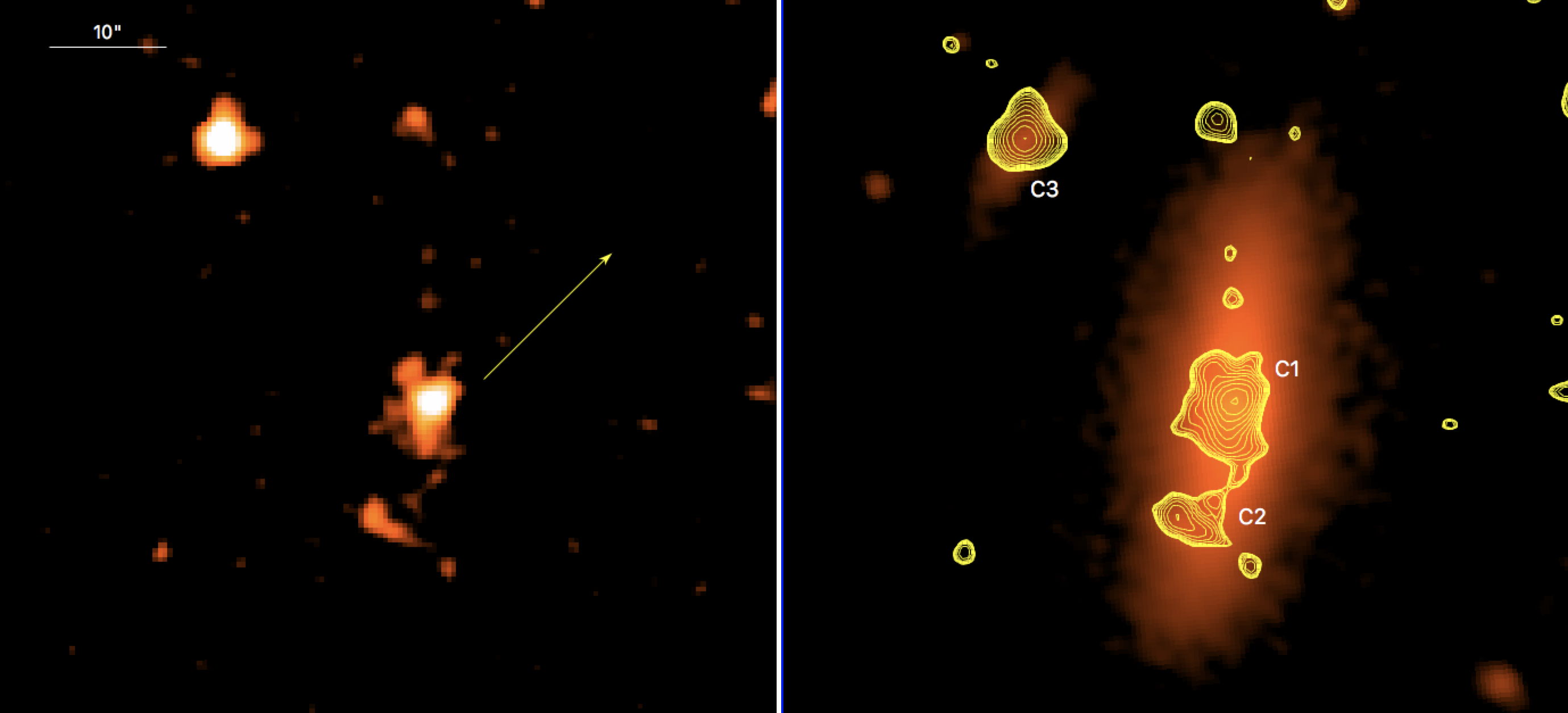}
  \includegraphics[scale=0.346]{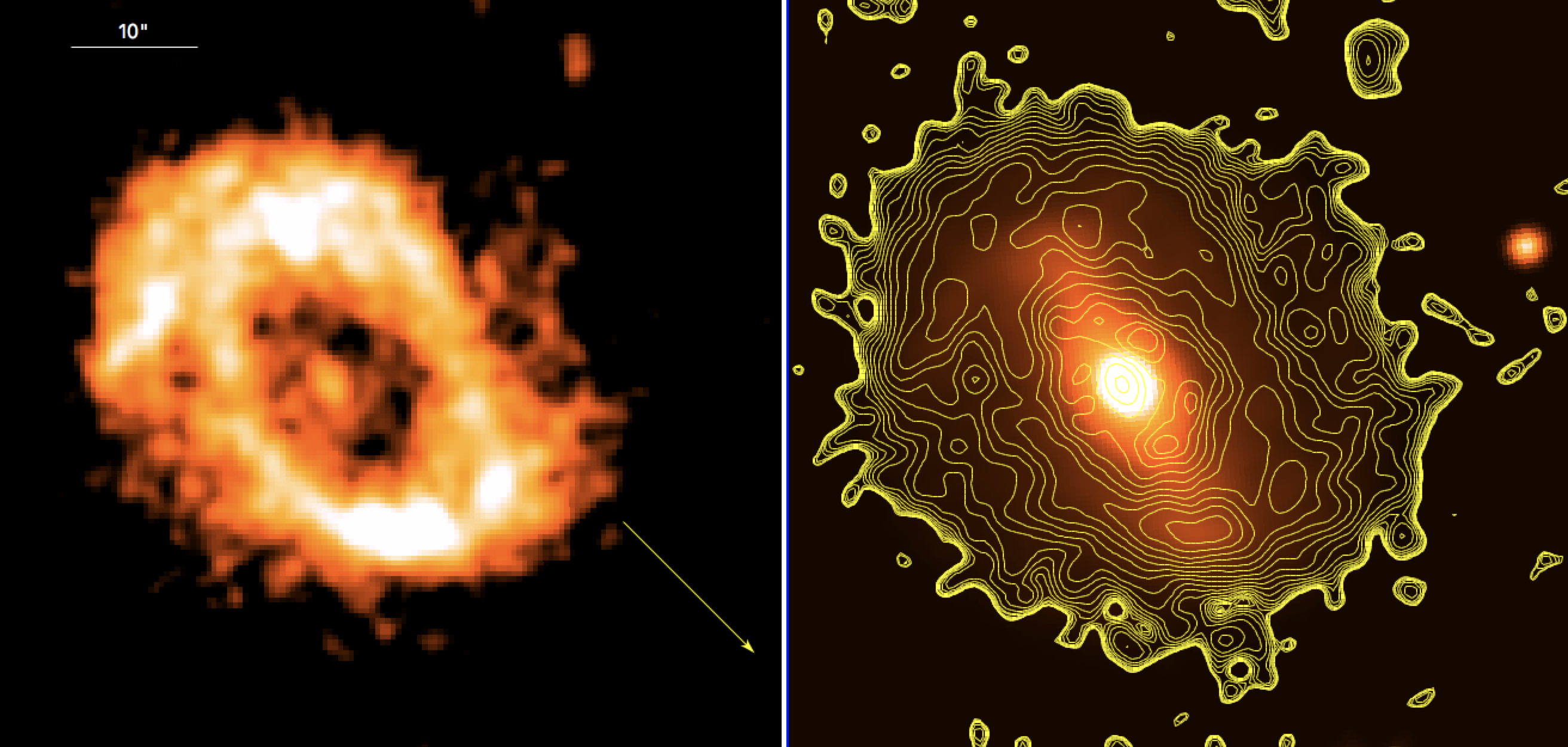}
  \includegraphics[scale=0.329]{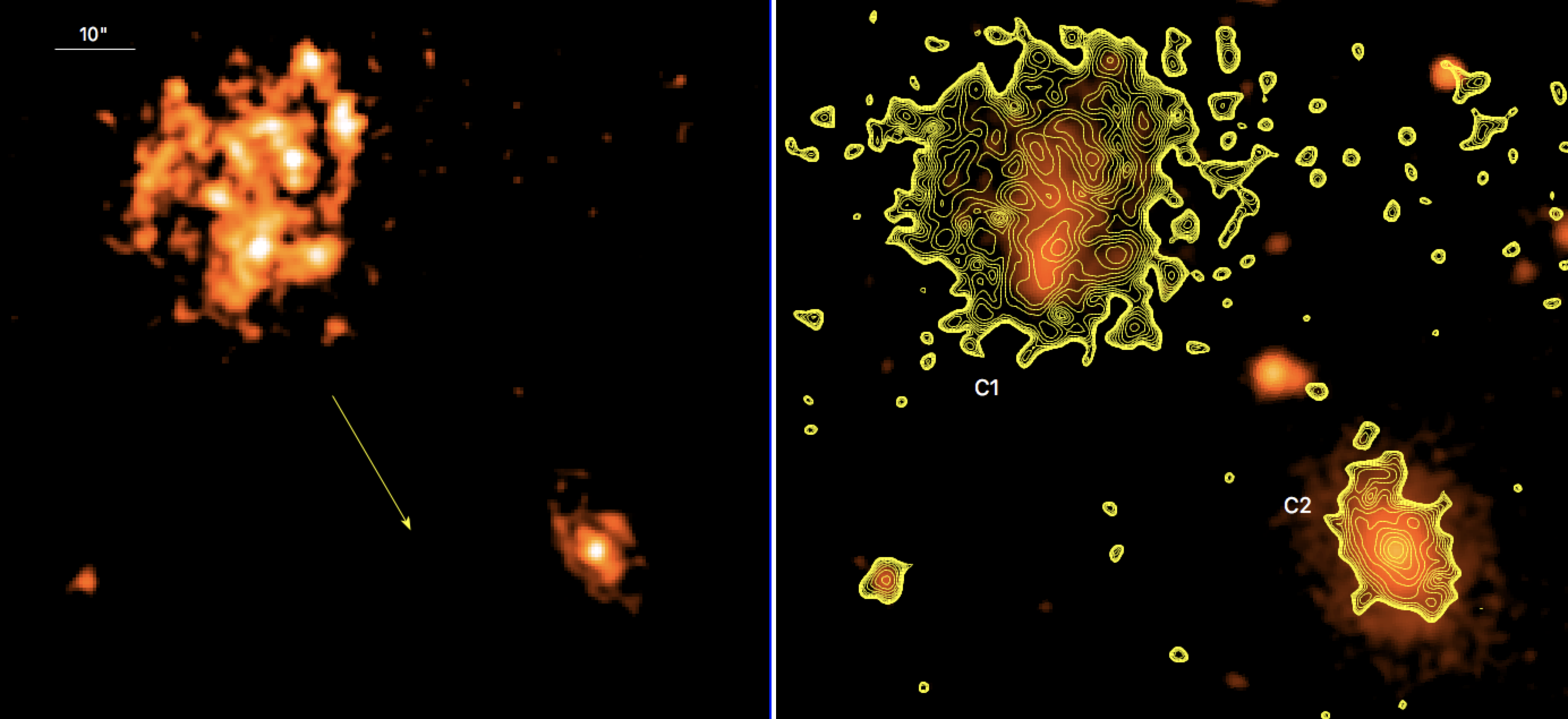}
  \caption{Same as Fig.~\ref{images}, but for (a) GMP 2347 , (b) NGC 4907, and (c) GMP 2943 and GMP 2989.   }
 \label{images4}
 \end{figure*}

 GMP 2943 (C1 in Fig.~\ref{images4}(c)) is a distorted spiral galaxy in the Coma cluster, whose optically-red nuclear region is observed below the blue spiral arms which take on a spider-like morphology in the optical image as shown in Fig.~\ref{g2943}. However, in the {\it FUV} image this galaxy appears to be a face-on spiral, much unlike it's optical facade.
 
 GMP 2989, which is also a Coma cluster member, is also seen alongside GMP 2943 in the image (Fig.~\ref{images4}(c)) as C2. This spiral galaxy has it's major axis pointed along the vector pointing in the direction of the cluster centre. GMP 2943 and 2989 are $\sim 1380$ km s$^{-1}$ away in redshift, hence unlikely to be interacting with each other, but within $\sim 0.45$ Mpc from the cluster centre. GMP 2989 is also the only bonafide AGN host galaxy (see Fig.~\ref{bpt}) amongst the galaxies with unusual {\it FUV} morphology discussed in this section.  
 
 \subsection{GMP~3317}
 GMP 3317 appears as a small, red, faint spheroid in the optical image, very close to a bright star. Not much is known about this distant galaxy ($z=0.2748$). However, the {\it FUV} image inevitably shows that this is likely a much larger galaxy, only whose nucleus is observed in the optical images. Interestingly though, there appears to be a hole in the {\it FUV} emission surrounding the nuclear region as seen in Fig.~\ref{images5}(a).
 
 \subsection{GMP~3170}
\begin{figure*}
  \includegraphics[scale=0.339]{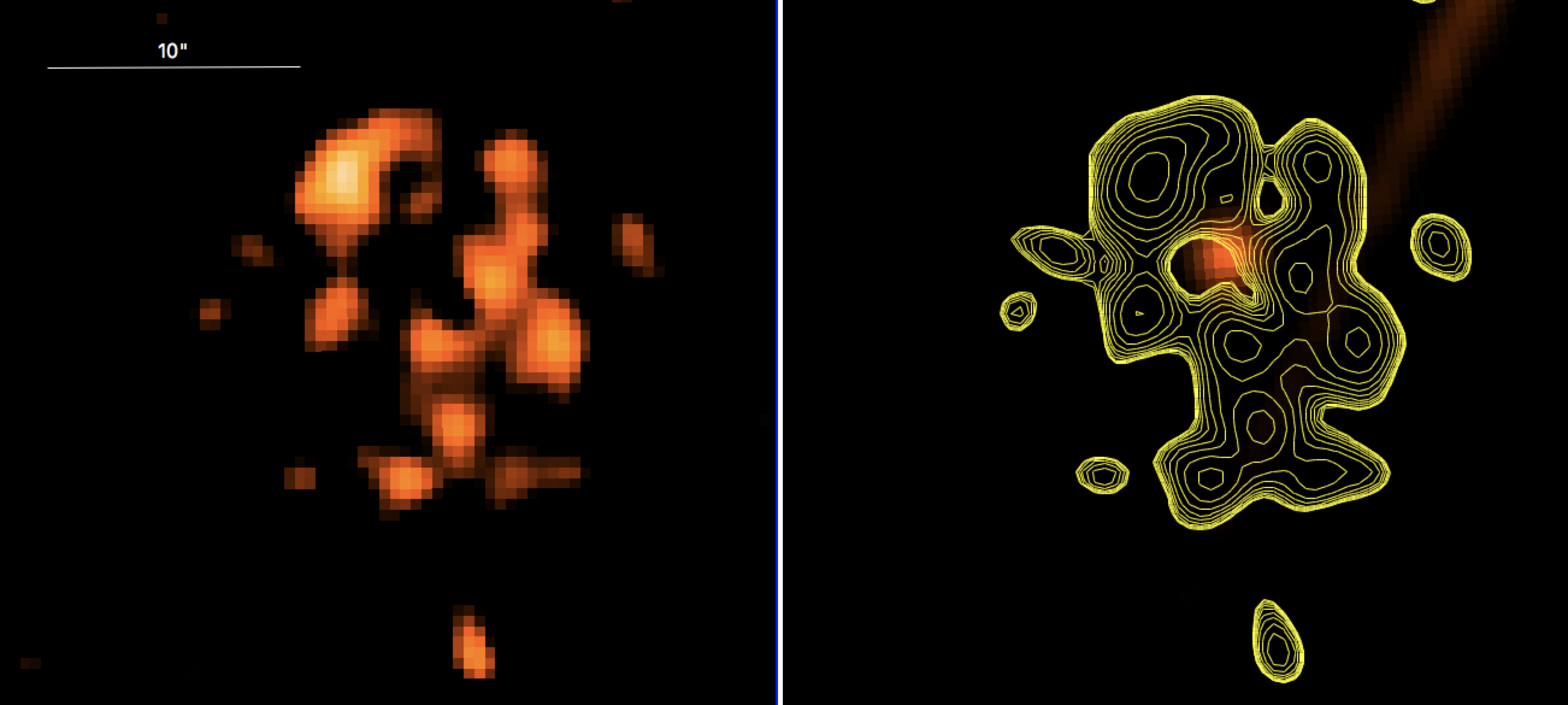}
  \includegraphics[scale=0.34]{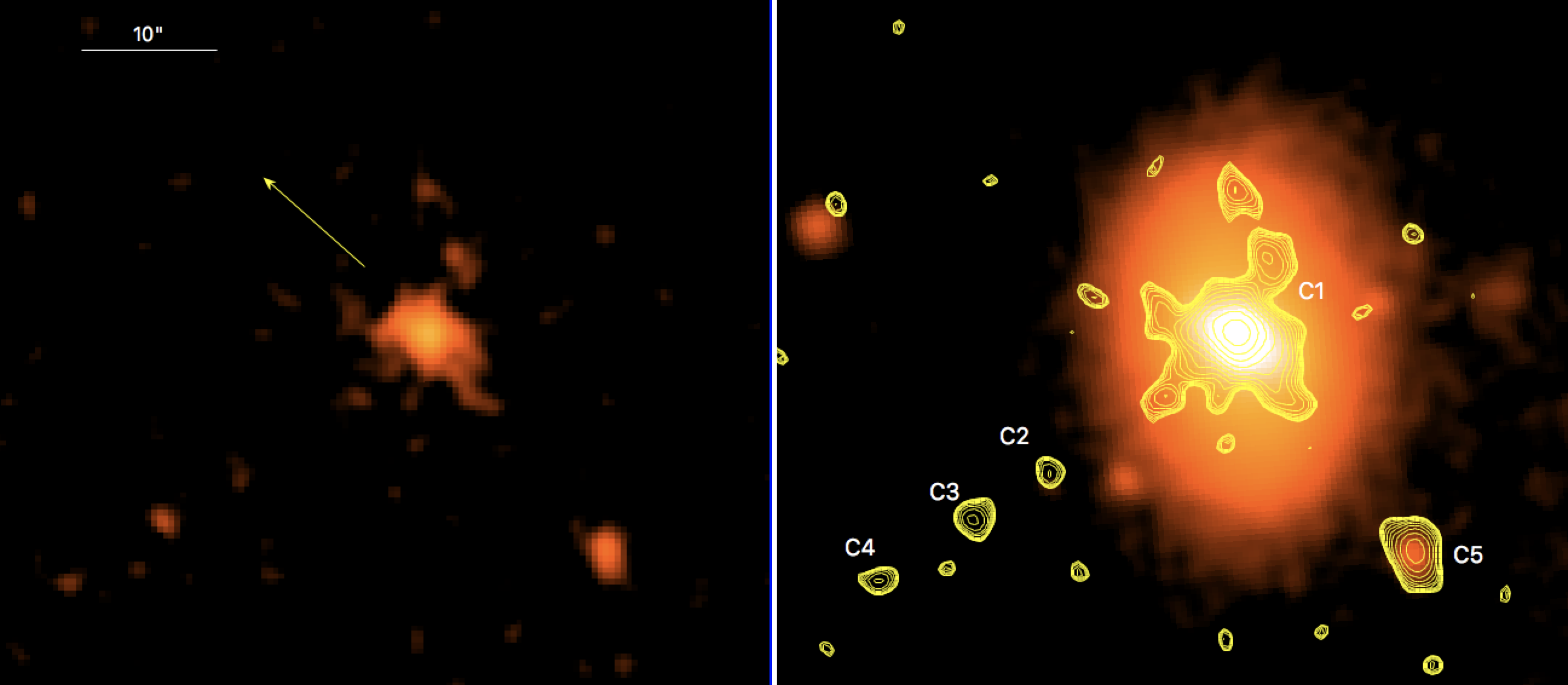}
  \includegraphics[scale=0.354]{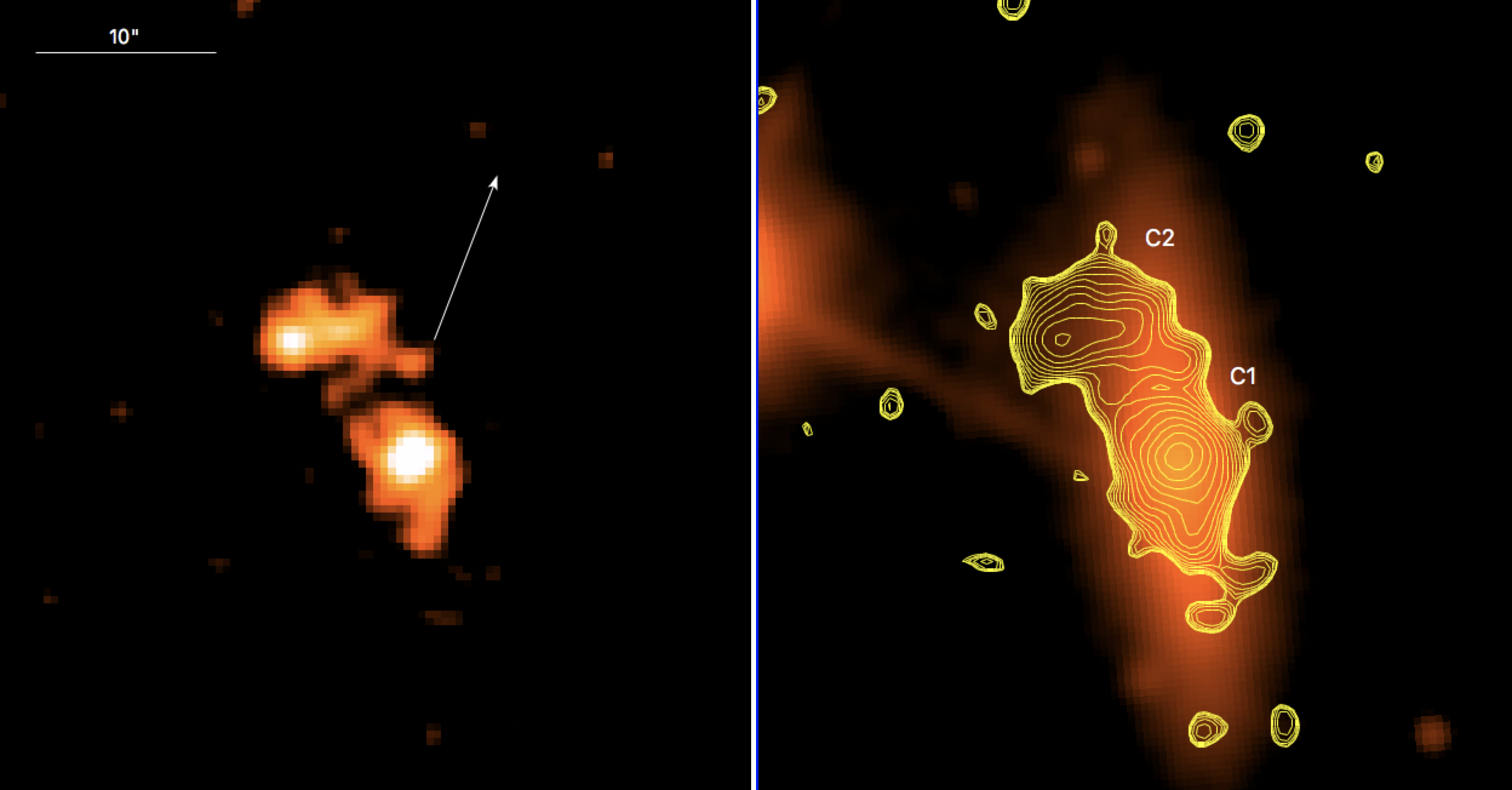}
  \caption{Same as Fig.~\ref{images}, but for (a) GMP 3317, (b) GMP 3170, and (c) GMP 2956.   }
 \label{images5}
 \end{figure*}

\begin{figure}
  \includegraphics[scale=0.23]{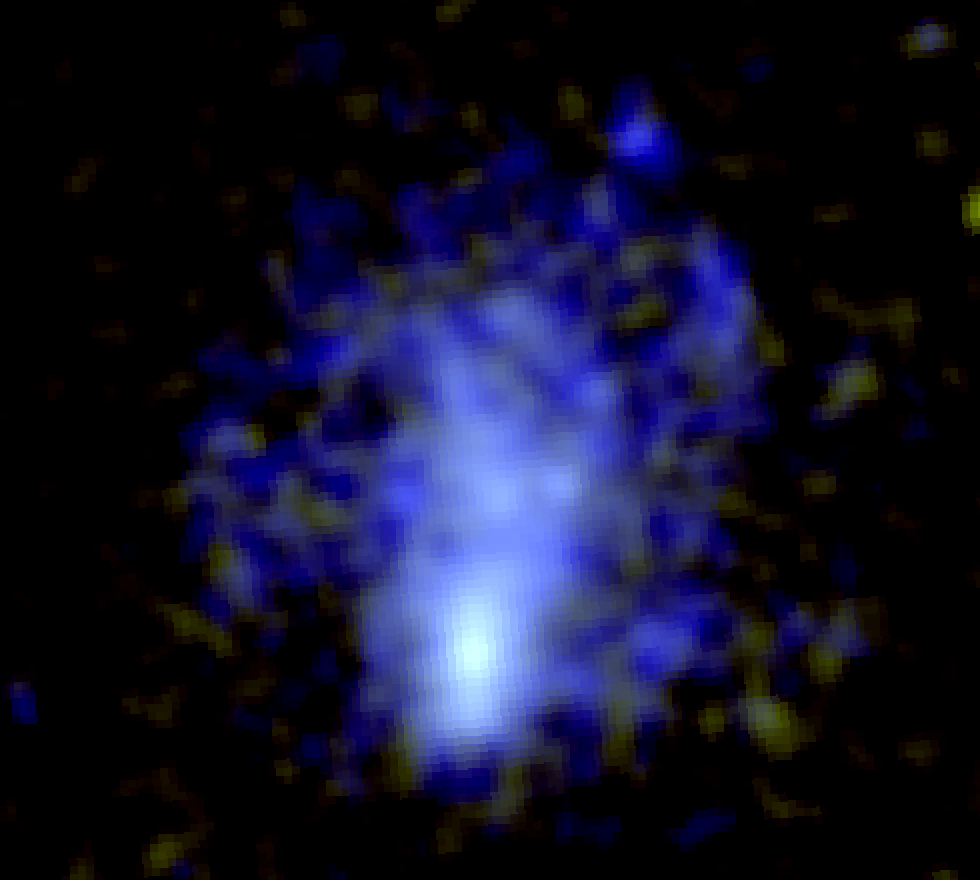}
  \vspace{0.1cm}
  \includegraphics[scale=0.282]{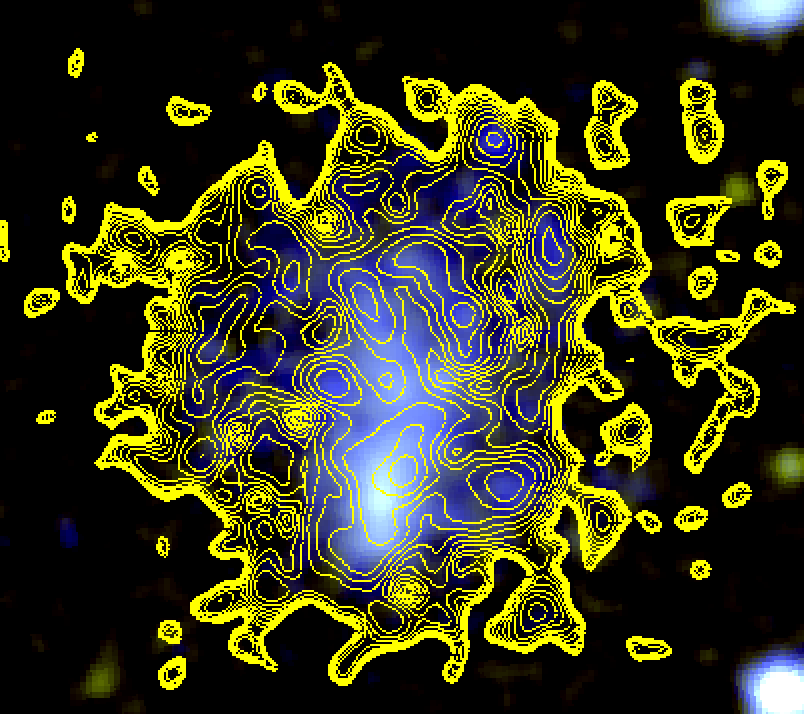}
    \caption{This figure shows the SDSS {\it gri} image of the galaxy GMP~ 2943 {\it (left)}, and the same with UVIT {\it FUV} contours overlaid  {\it (right)}. Also, see 
    Fig.~\ref{images4}(c).   }
 \label{g2943}
 \end{figure}

 This Coma cluster galaxy is only $\sim 0.04$ Mpc from the cluster centre, and hence very likely experiencing the hostile environmental mechanisms. Besides the central component C1 in Fig.~\ref{images5}(b), one can clearly see a tail of {\it FUV} blobs (components C2, C3 and C4) in the direction perpendicular to the vector pointing towards the cluster centre. Although follow-up spectroscopic observations are required to confirm our speculation that the blobs are connected to GMP 3170.
 
 \subsection{GMP~2956}
 This S0 Coma cluster galaxy lies at a distance of $\sim 0.31$ Mpc from the cluster centre. The {\it FUV} emission appears to be divided into at least two components: C1 is centered at the core of the galaxy, while the component C2 does not show any obvious optical counterpart. C2 also appears to be tilting in the north-east direction away from the direction of the cluster centre. 
 
 \subsection{Optically-faint candidates}
\begin{figure}
  \includegraphics[scale=0.171]{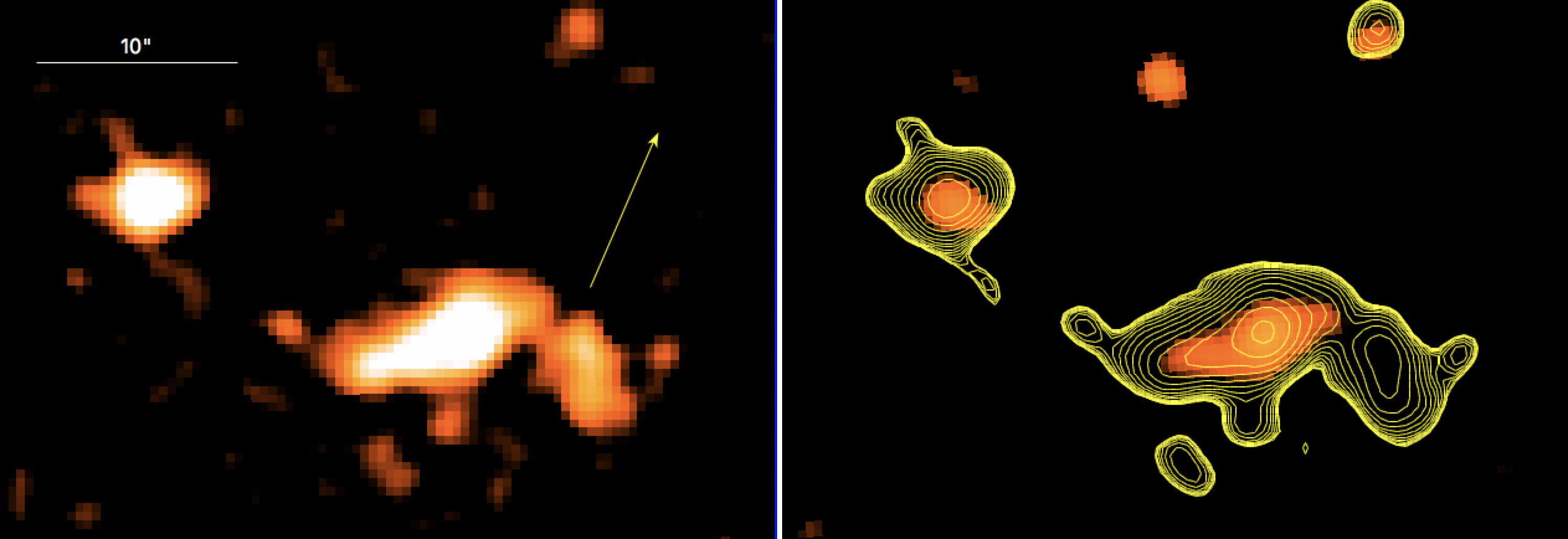}
  \includegraphics[scale=0.171]{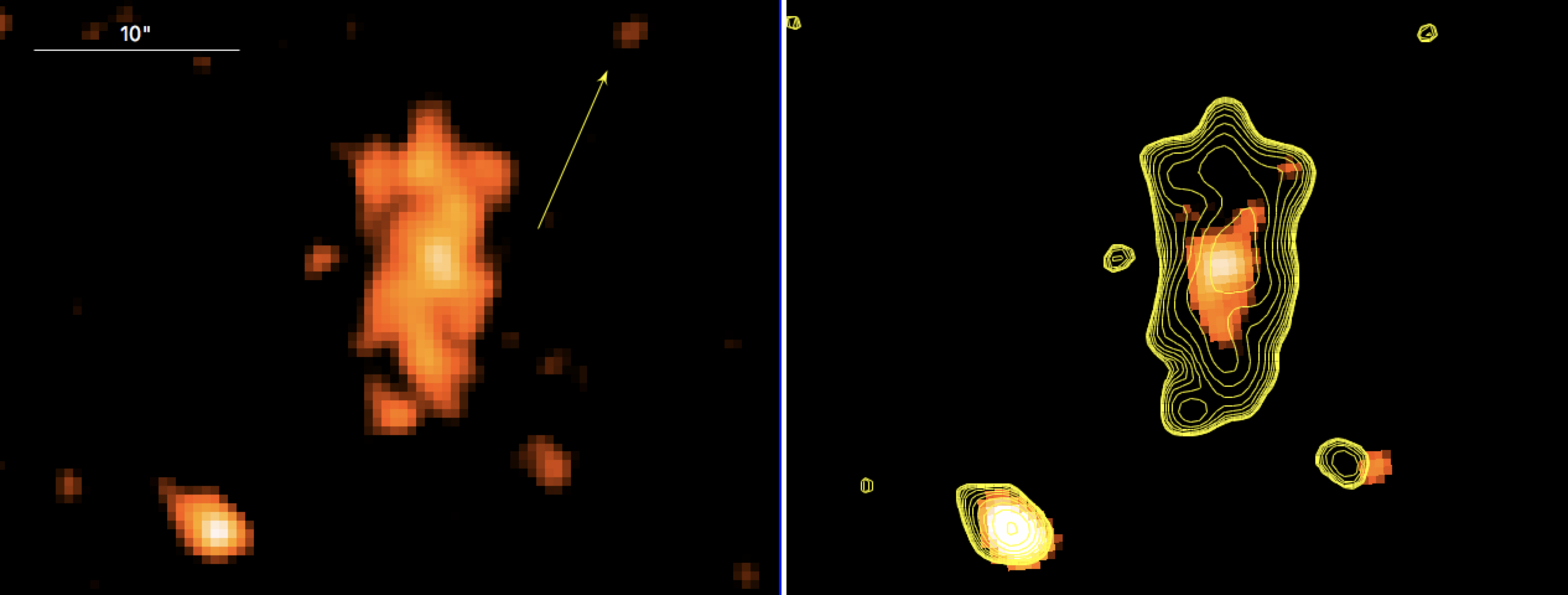}
  \includegraphics[scale=0.164]{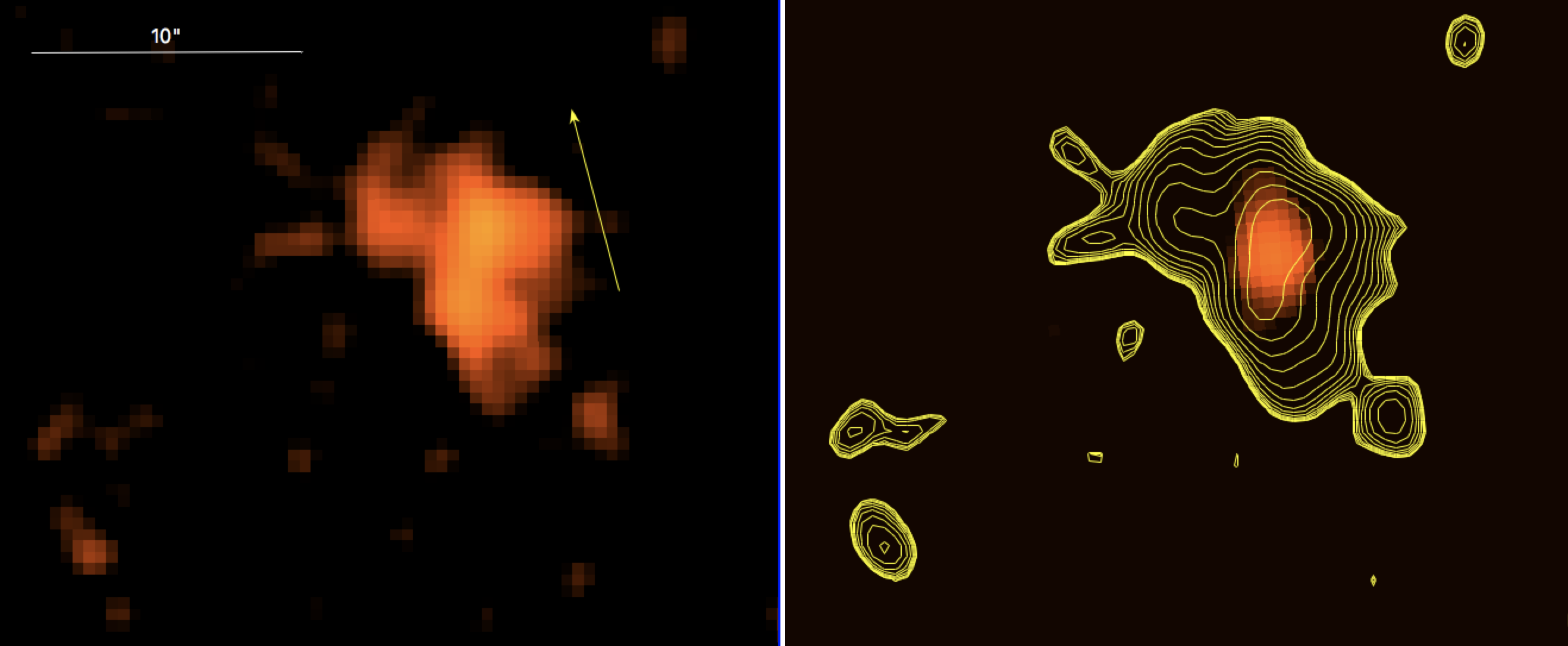}
  \includegraphics[scale=0.156]{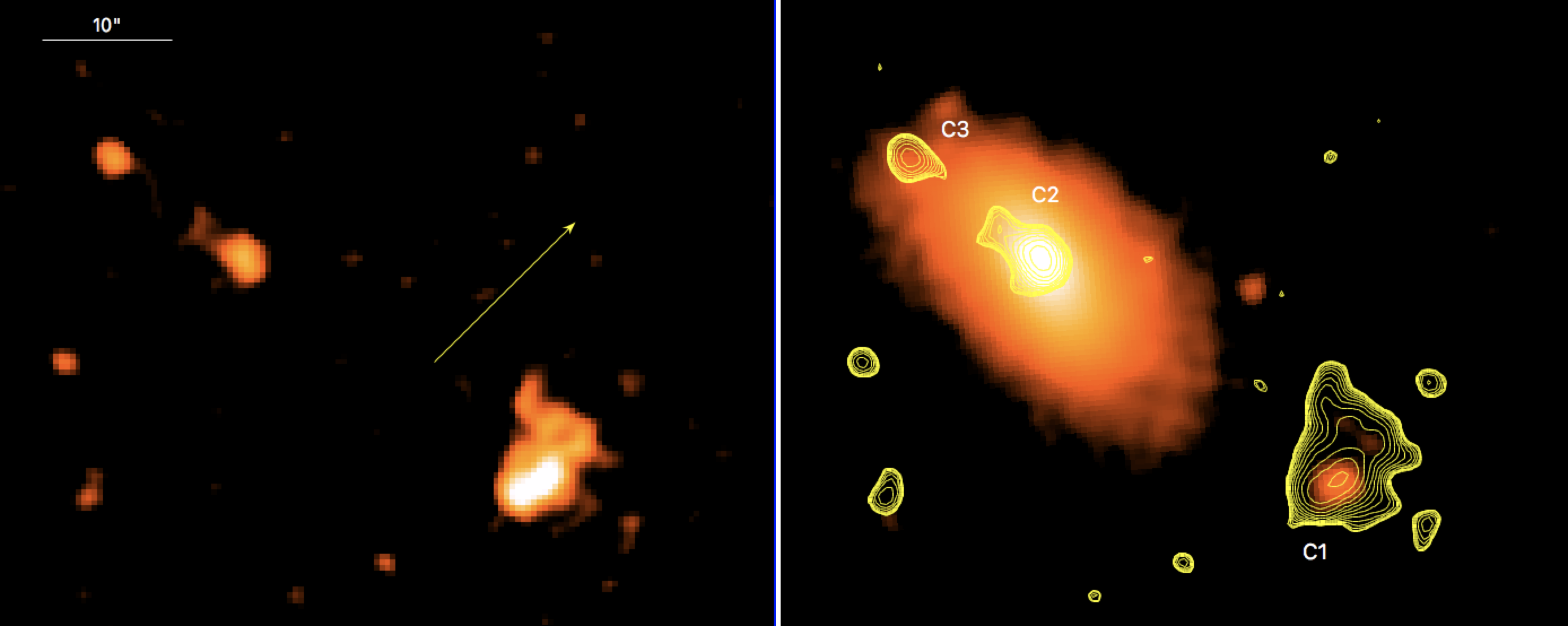}
  \includegraphics[scale=0.198]{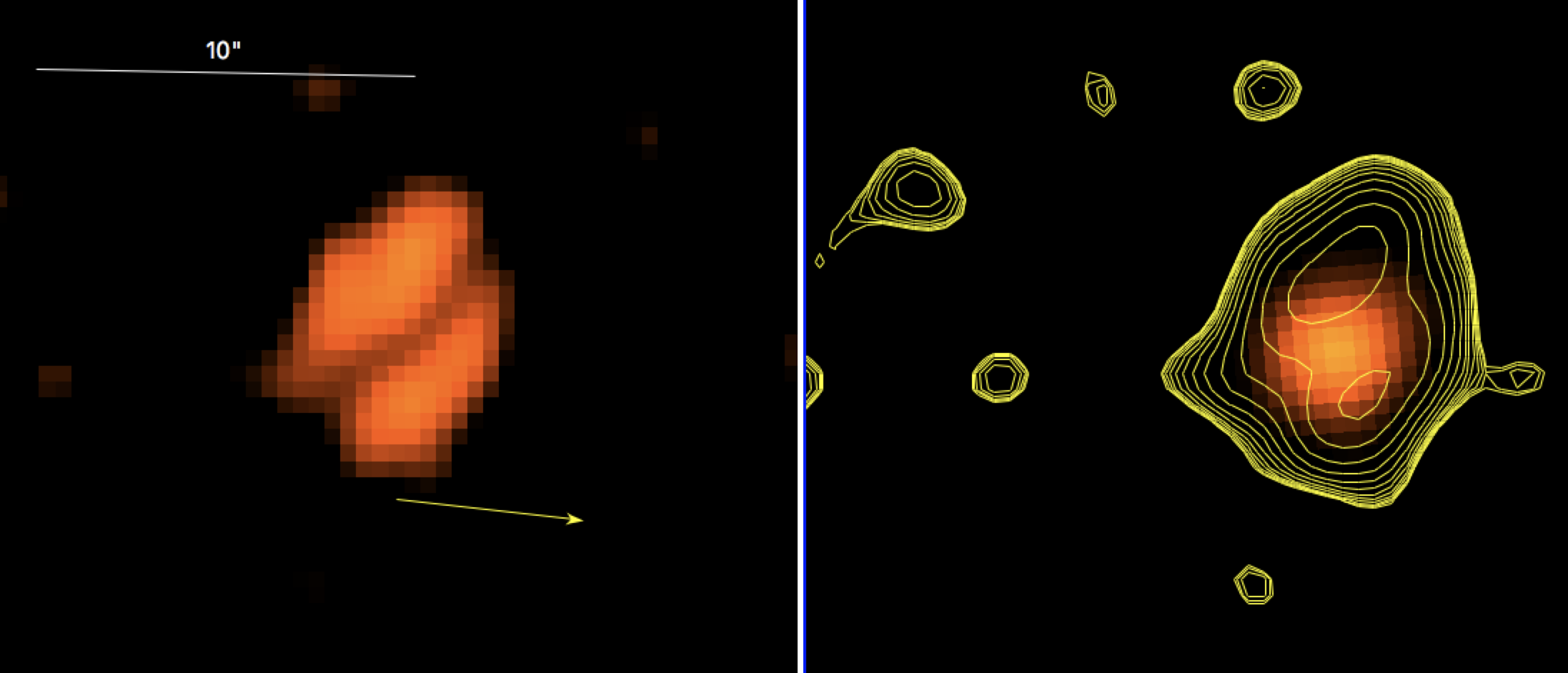}
  \includegraphics[scale=0.168]{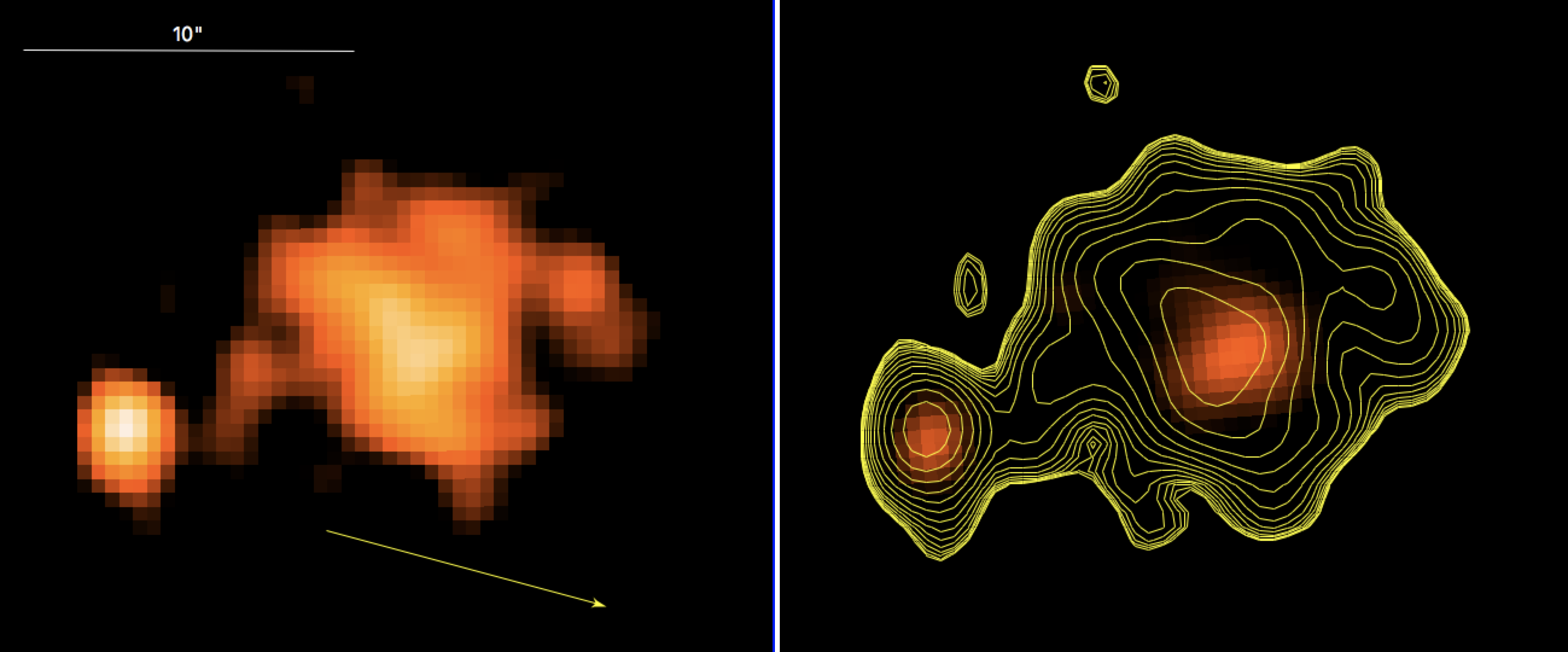}
  \includegraphics[scale=0.17]{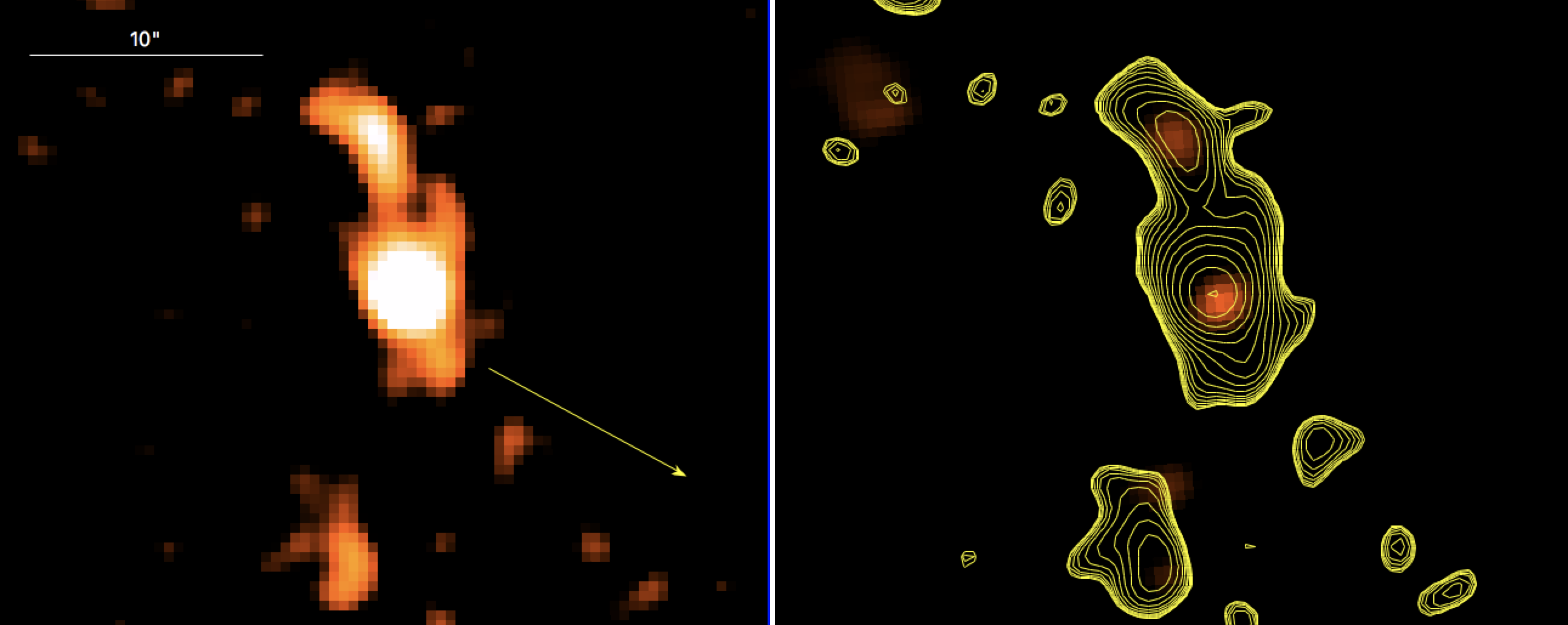}
  \caption{Same as Fig.~\ref{images}, but for {\it (from top to bottom:)} (a) GMP 2717, (b) Subaru ultra-diffuse galaxy, (c) Galaxy1, (d) GMP 2476, (e) GMP 2454, (f) GMP 2320 and (g) Galaxy2.   }
 \label{images6}
 \end{figure}

 Besides the above mentioned galaxies, we also detected a few optically faint galaxies for which redshift information is not available, but their {\it FUV} morphology is markedly different from the optical image. These galaxies are listed at the bottom of Table~\ref{tab:HI}, shown in Fig.~\ref{images6}, and are briefly described below.
 \begin{itemize}
   \item {\it GMP~2717:} This faint irregular galaxy shows an extended emission pointing opposite to the cluster centre on the southwestern side of the image.   
   
   \item {\it Subaru-UDG:} This ultra-diffuse galaxy (UDG) was first reported by \citet{yagi16} in their Subaru imaging survey of UDGs in the Coma cluster region. The {\it FUV} extent of the galaxy is evidently larger than it's optical size. 
   
   \item {\it Galaxy 1:} This is also an irregular low surface brightness galaxy, which is much larger in {\it FUV} relative to the optical extent.
   
   \item {\it GMP~2476:} This is component C1 in Fig.~\ref{images6} (d), and like other faint dwarf galaxies, appears to be much smaller in the optical image. Also seen in the field GMP~2457 which shows nuclear {\it FUV} emission (component C2) and an off-centre component C3. The entire system is aligned normal to the vector pointing in the direction of the cluster centre. 
   
   \item {\it GMP~2454:} This faint dwarf galaxy appears as a small, red spheroid in the multi-band SDSS data, but shows a distinct double-object morphology in the {\it FUV} image.   
   
   \item {\it GMP~2320:} Only the nuclear region of this dwarf galaxy seems to be observable in the optical image, while the larger extent is apparent in the {\it FUV} data. The point source in the southeastern corner may be a star unrelated to the galaxy. 
   
   \item {\it Galaxy 2:} This dwarf galaxy shows a very interesting morphology in the {\it FUV} image, but interestingly appears as two separate faint spheroidal objects in the optical data. A distinct tail of {\it FUV} emitting blobs can also be observed extending towards southwest of the field. 
 \end{itemize}
 
 In summary, at least 13 of the 23 galaxies discussed in this section are Coma cluster members likely to be experiencing ram-pressure stripping. Although gravitational influence of the neighbouring galaxies may also have contributed to their unusual {\it FUV} morphology, it is worth noting that all of them are $\lesssim 0.5 h^{-1}$ Mpc of the centre of the Coma cluster. The seven low surface brightness dwarf galaxies will be explored further in future work.   

 \section{Discussion}
 \label{discuss}
 
 The aim of this paper is to analyse the sources detected in our deep {\it FUV} UVIT data for the central region of the Coma cluster. In order to do so, we have employed the 
 {\it NUV} and {\it FUV} data from the \g mission, and optical photometric and spectroscopic data from the SDSS as well. We have presented various statistical properties of the sources detected in our data, classifying them into Coma cluster members, other galaxies, stars and quasars, respectively, where possible. In the following, we compare 
 our data and findings with the existing literature, and further discuss the properties of the galaxies having unusual morphology in the {\it FUV} band. 
 
 \subsection{Phase space distribution in Coma cluster}
 
 \begin{figure}
 {\includegraphics[scale=0.32]{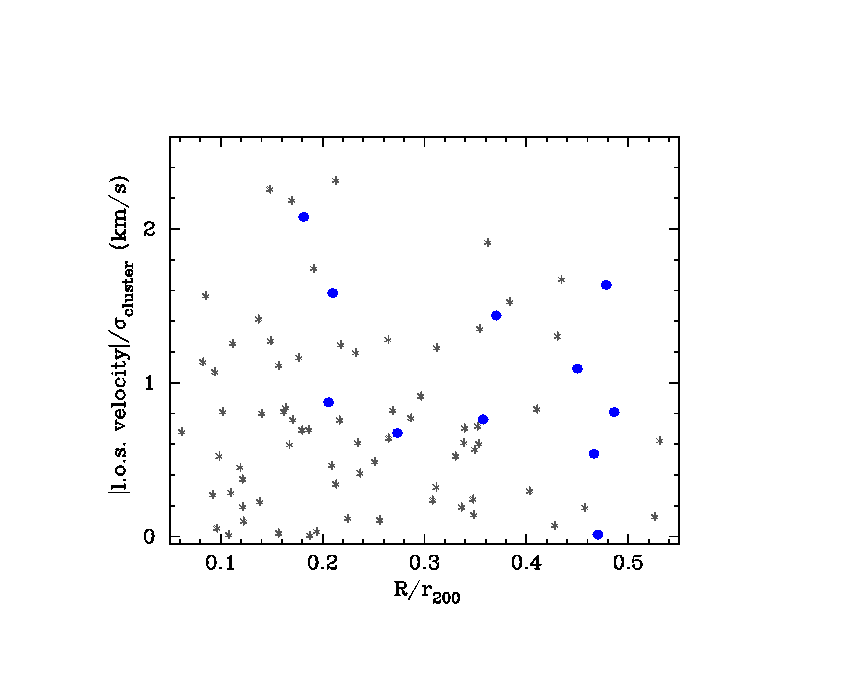}}
 \caption{This figure shows the distribution of all galaxies ({\it grey asterisks}) and those with distorted UV morphology ({\it blue points}) in the 
 phase space around the Coma cluster centre. It is evident that most of the UV galaxies lie away from the cluster core in spatial as well as velocity space. } 
 \label{rv}
 \end{figure}
 
 The phase space diagram mapped by the distance of a galaxy to the cluster centre normalized by $r_{200}$\footnote{$r_{200}$ is the radius at which the mean interior 
 over-density in a sphere of radius $r$ is 200 times the critical density of the Universe.} and the line-of-sight velocity of the galaxy scaled by the velocity 
 dispersion\footnote{Although there are multiple values of $\sigma_v$ present in the literature, for simplicity we assume $\sigma_v = 1000$ km s$^{-1}$ here.} is 
 shown in Fig.~\ref{rv}. This figure shows that most of the Coma cluster galaxies having distorted UV morphologies are far away from the cluster centre, in 
 spatial as well as velocity space. This suggests that these galaxies have recently entered the cluster, and are experiencing the impact of the intra-cluster 
 medium (ICM) for the first time. 
  
 \subsection{Spectroscopic properties of UVIT sources}
 
  \begin{figure}
 {\includegraphics[scale=0.29]{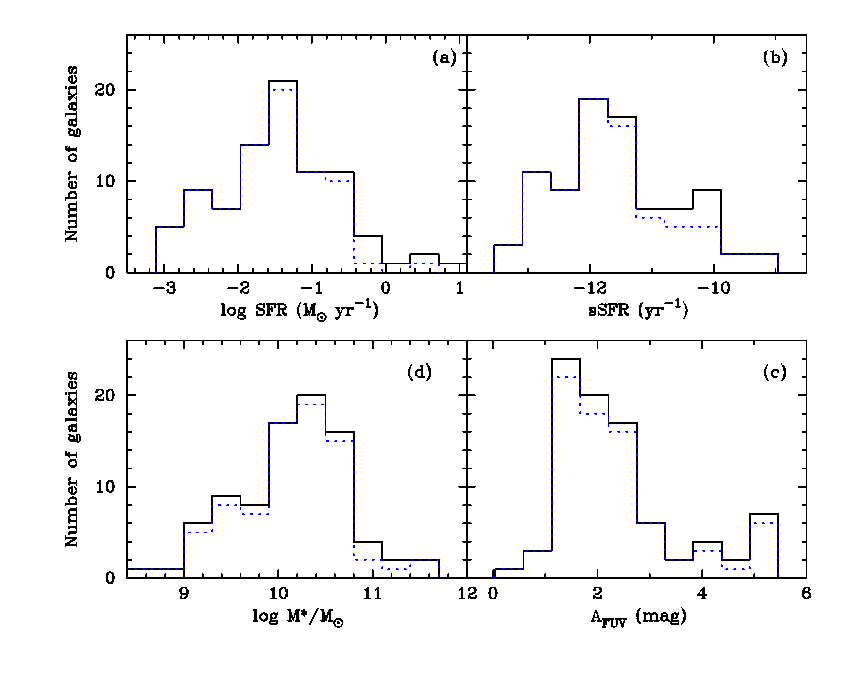}}
 \caption{This figure shows the distribution of (a) log SFR, (b) sSFR, (c) log $M^*$ and (d) $A_{FUV}$ for the galaxies matched to our UVIT sources in the GSWLC database. 
 The {\it solid black lines} are for all the sources, while the {dotted blue line} represents the Coma cluster members, respectively.  } 
 \label{gswlc}
 \end{figure} 
 
 We found 88 galaxies from our UVIT sample in the {\it GALEX-SDSS-WISE} Legacy catalogue \citep[GSWLC henceforth;][]{salim16}, 79 of which are 
 members of the Coma cluster. The GSWLC provides estimates of physical properties for $\sim 700,000$ galaxies in the {\it Galex} footprint, also observed by the SDSS. 
 The physical properties of galaxies are obtained by fitting the UV and optical spectral energy distribution following a Bayesian methodology. In Fig.\ref{gswlc} we show 
 the distributions of star formation rate (SFR), stellar mass ($M^*$), SFR/$M^*$ and extinction in the FUV band ($A_{FUV}$), for all the matched 
 galaxies and the Coma cluster members, respectively.  
  
 \begin{figure}
 {\includegraphics[scale=0.33]{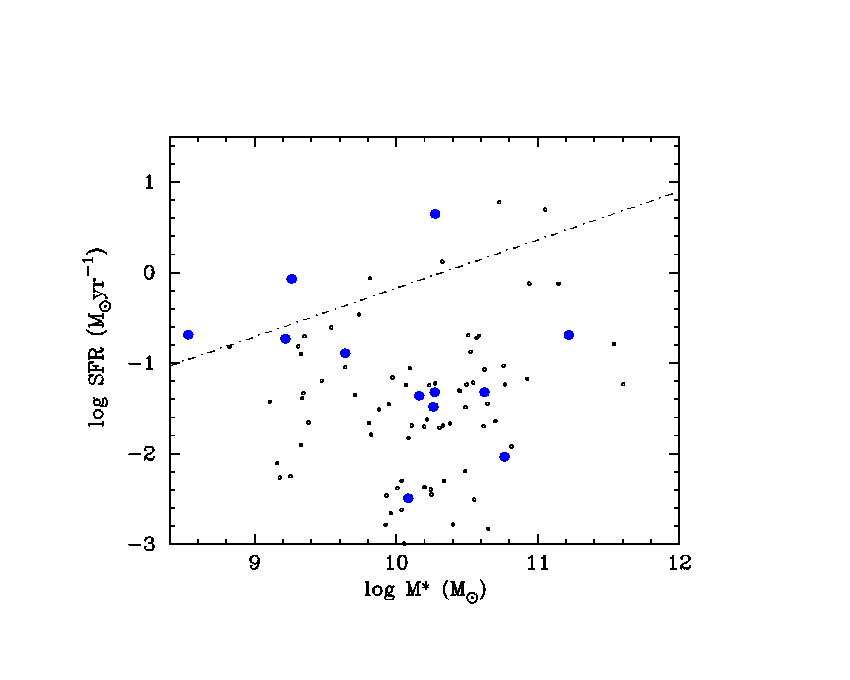}}
 \caption{This figure shows the distribution of all UVIT detected galaxies found in the GSWLC catalogue ({\it black points}), and Coma cluster galaxies with unusual morphology ({\it blue asterisks}) in the stellar mass--SFR space. The dot-dashed line represents the main sequence of star-forming field galaxies \citep{roberts20}. While three distorted 
 galaxies lie $\lesssim 0.5$ dex away from the main sequence, in the region where ram-pressure stripping candidates are expected to lie \citep[see fig.~7 of][]{roberts20}, 
 majority of the galaxies with unusual morphology populate the region of this space expected to be inhabited by the red sequence and normal star-forming galaxies of the Coma cluster.   } 
 \label{sf-mass}
 \end{figure} 

 In Fig.~\ref{sf-mass} we show the SFR of the matched GSWLC galaxies as a function of their stellar mass. It is well known that the SFR of galaxies correlate well with their 
 stellar mass, such that more massive galaxies have higher SFR relative to their lower mass counterparts. In a recent 
 work, \citet{roberts20} have used a bayesian-based regression approach on the SDSS data release 12 data to find the main-sequence relationship for star-forming field 
 galaxies as $\rm{log~SFR} = 0.55\times {\rm log}~M^*-5.7$. This relation\footnote{The stellar masses and SFR for their 
 work are in turn taken from the GSWLC-2 SED fitting catalogue \citep{salim16, salim18}.} is shown for reference as a dotted line in Fig.~\ref{sf-mass}, along with the 
 UVIT galaxies, and galaxies with unusual {\it FUV}  
 morphology found in the GSWLC catalogue. Most of our UVIT detected galaxies are found in the region occupied by the passive galaxies 
 (log sSFR$< -11$ yr$^{-1}$; where, sSFR $\equiv$ SFR/$M^*$) in the SFR-$M^*$ space, evidently showing that star formation activity is suppressed in the cluster environment, 
 even for UV-bright galaxies. We also note that nine of the twelve galaxies with unusual {\it FUV} morphology are found below the field main sequence 
 relation, in the region occupied by passive galaxies in the sample of \citet{roberts20}. These authors however, found their ram-pressure stripping candidates to 
 lie above the main sequence, where only three of our distorted galaxies make an appearance \citep[][see their fig.~7]{roberts20}. 
 This analysis suggests that beside ram-pressure stripping galaxies, unusual {\it FUV} morphology may be found among transitional galaxy populations. 
 We also note that the low-surface brightness dwarf galaxies do not make an appearance on this plot due to lack of spectroscopic data. 
 
 Recently, \citet{chen20} presented high spatial resolution HI and deep 1.4GHz continuum data for 20 galaxies which are being ram-pressure stripped (RPS) in the Coma cluster. 
 Three of these galaxies (GMP 2910, GMP 2559 and GMP 3016) are also part of our UVIT data. While GMP 2910 and GMP 2559 exhibit RPS tails matching their 
 H$\alpha$ observations \citep{smith10, yagi10}, GMP 3016 does not show RPS tails in the radio continuum. 

 \begin{figure}
 \vspace{-15mm}
 \hspace{-5mm}
 \vspace{-10mm}
 {\includegraphics[scale=0.35]{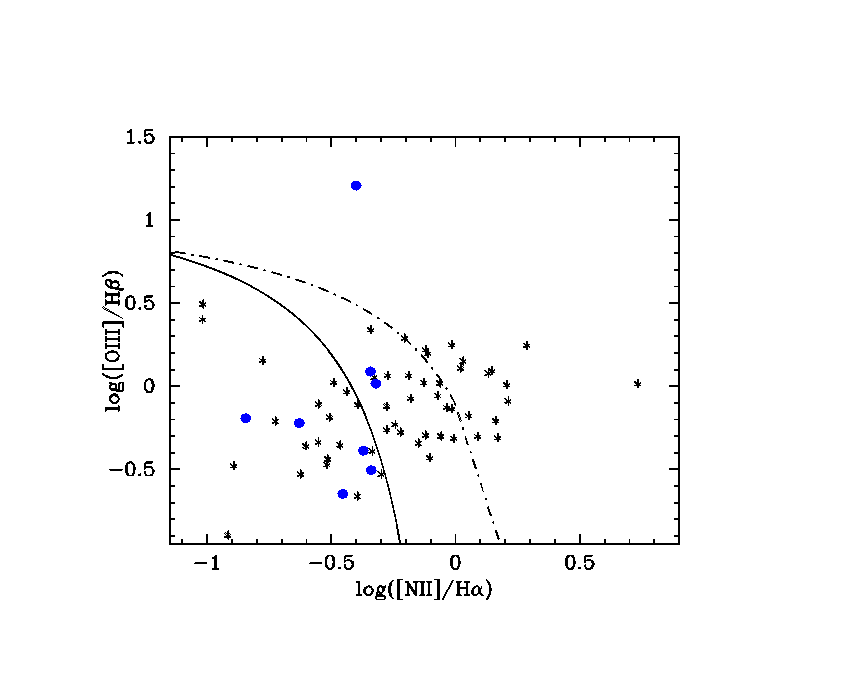}}
 \caption{This figure shows the distribution of 48 UVIT detected galaxies found in the GSWLC catalogue ({\it grey asterisks}), and those with distorted morphology ({\it blue
 points}) in the BPT plane. The {\it solid} and {\it dot-dashed} lines mark the limiting criteria to distinguish between star-forming and AGN galaxies, respectively. The galaxies 
 inhabiting the region between these two lines are known as composites, i.e. part of their emission comes from an AGN-like component. We note that 18 and 17 galaxies
 respectively classify as star-forming and composites, while the remaining 13 as AGN. While amongst the galaxies with distorted UV morphology, five are star-forming, two are composite and one is an AGN. } 
 \label{bpt}
 \end{figure} 

 There are 48 galaxies matched with the GSWLC, and having good signal-to-noise in all four emission lines (H$\alpha$, [NII], [OIII] and H$\beta$) in the SDSS databse, which allows us to distinguish 
 the active galactic nuclei (AGN) from the star-forming galaxies using the \citet{bpt} criteria. The Baldwin, Phillip, Terlevich, popularly known as the BPT diagram is shown in Fig.~\ref{bpt}, with the galaxies with distorted UV 
 morphology highlighted where possible. This space can be further divided into the regions occupied by star-forming, AGN and composites (i.e. galaxies with a weak AGN-like
 emission component) using the well used criteria put forward by \citet{kauff03} and \citet{kewley01}. As shown in Fig.~\ref{bpt}, we find that 18 UVIT sources with optical spectra are star-forming galaxies, 17 are composites and the remaining 13 are AGN host galaxies. Amongst the galaxies with unusual UV morphology, five are star-forming galaxies, two are found in the region occupied by the composite galaxies, and one is an AGN. Since star-forming galaxies are in general comparable in number to the composite galaxies, it would be interesting to explore the {\it FUV} emission in AGN host galaxies for a similar, larger sample in future. 
 
 We conclude this discussion by summarising some of the limitations of the data analysed in this paper. Our UVIT data are limited to the {\it FUV} band only. 
 The {\it NUV} band was not operated due to the presence of bright point sources in the field. Instead, we have used the {\it NUV} data from {\it GALEX}.
 The large difference in the resolution and the sensitivity of the two instruments, however, made a comparison of the {\it FUV} and {\it NUV} emission from the galaxies difficult. 
 We wish to point out that since this is the first time that this field has been observed and analysed to such a depth in the {\it FUV} waveband, many interesting sources 
 detected in the UVIT field lack redshift information, and require follow-up optical observations to throw more light on their origin and properties.

\section{Summary}
\label{summary}
 
 In this paper we have examined a deep {\it FUV} image of the central region of the Coma cluster, observed by the UVIT onboard the Indian multi-wavelength satellite 
 mission {\it AstroSat}. For the sources detected by both missions, we find a good correlation between the flux derived from the UVIT BaF$_2$ filter and the two \g wavebands. 
 The UVIT fluxes of the detected sources (galaxies, stars and 
 quasars) span two orders of magnitude in flux. We find that most of the galaxies brighter than $r \sim 17$ mag are members of the 
 Coma cluster, and occupy the bright end of the distributions in all colour-magnitude and, red end in the colour-colour space, respectively. Amongst others, we have also detected three 
 quasars, one of which at $z=2.315$ is likely the farthest object observed by the UVIT so far. 
   
 Furthermore, we investigated sources with unusual {\it FUV} morphology, twenty three of which are discussed in detail here. Amongst others, we have identified several new candidate galaxies having unusual $FUV$ morphology, many of which could be members of the Coma cluster. New, spectroscopic data are however needed to confirm this hypothesis.  Our analysis indicates that distorted FUV sources may have recently entered the Coma cluster, and hence undergoing stripping events under the influence of the cluster-related environmental mechanisms. While five of these galaxies could be classified as star-forming on the basis of their emission line properties, two are found to be composites and one is an AGN.
  
 To the best of our knowledge, this is the first study of a galaxy cluster field being carried out with the UVIT data. Hence, despite several limitations mentioned 
 in the previous section, this work provides an idea of the finesse with which UVIT has captured the features of nearby extra-galactic objects, and a flavour of the problems which, together with the multi-band archival data may be addressed using the UVIT. In order to facilitate this, we have also provided detailed information for data handling and reduction for fellow UVIT users who intend to work with similar data.   
  
\section{Acknowledgements}
 
 S. Mahajan was funded by the INSPIRE Faculty award (DST/INSPIRE/04/2015/002311) and the SERB Research Scientist (SRS) award (SB-SRS/2020-21/56/PS), 
 Department of Science and Technology (DST), Government of India. 
 K. P. Singh thanks the Indian National Science Academy for support under the INSA Senior Scientist Programme. Authors are very grateful to Prof. S. Tandon for his
 valuable comments on earlier versions of this manuscript. 
 We are very grateful to the anonymous reviewer for their constructive critique which helped in improving this manuscript. 
 
 This publication uses data from the {\it AstroSat} mission of the Indian Space Research Organisation (ISRO), archived at the Indian Space
 Science Data Centre (ISSDC). UVIT project is a result of collaboration between IIA (Bengaluru), IUCAA (Pune), TIFR (Mumbai), several  
 centres of ISRO, and the Canadian Space Agency (CSA). 
 This research has made use of the NASA/IPAC Infrared Science Archive, which is funded by the National Aeronautics and Space Administration and 
 operated by the California Institute of Technology.
 The {\sc topcat} software \citep{taylor05} was used for some of the analysis presented in this paper.

\bibliographystyle{pasa-mnras}
\bibliography{coma-uvit}

\end{document}